\documentclass[aps,11pt,prd,groupedaddress,nofootinbib,notitlepage,eqsecnum,preprintnumbers]{revtex4-2}
\usepackage[utf8]{inputenc}
\usepackage{hyperref}
\usepackage{xcolor}
\usepackage{graphicx}
\usepackage{amsmath,amssymb}
\usepackage{bm}
\usepackage{comment}
\usepackage[shortlabels]{enumitem}
\usepackage{overpic} 
\usepackage{ulem}
\usepackage{makecell}
\usepackage{tabularx}
\usepackage{booktabs}
\usepackage{diagbox}
\usepackage{pifont}

\def\id{{\rm d}}

\begin{document}
\title{Regularizing the induced GW spectrum with dissipative effects}

\author{\textsc{Guillem Dom\`enech$^{a,b}$}}
    \email{{guillem.domenech}@{itp.uni-hannover.de}}

\author{\textsc{Jens Chluba$^{c}$}}
    \email{{jens.chluba}@{manchester.ac.uk}}

\affiliation{$^a$Institute for Theoretical Physics, Leibniz University Hannover, Appelstraße 2, 30167 Hannover, Germany.}
\affiliation{$^b$ Max-Planck-Institut für Gravitationsphysik, Albert-Einstein-Institut, 30167 Hannover, Germany}
\affiliation{$^c$ Jodrell Bank Centre for Astrophysics, School of Physics and Astronomy,\\
The University of Manchester, Manchester M13 9PL, U.K.}

\begin{abstract}
Finite mean free paths of light particles, like photons and neutrinos, lead to dissipative effects and damping of small-scale density fluctuations in the early universe. We study the impact of damping on the spectral density of gravitational waves induced by primordial fluctuations in the radiation-dominated universe. We show that the most important effects of damping are $(i)$ regularization of the resonant frequency and $(ii)$ a far low-frequency tail with no logarithmic running. The exact location of the break frequency below which the logarithmic running is lost depends on the damping rate. Both effects stem from the effective finite lifetime of the gravitational wave source caused by damping. Interestingly, we find that, for the standard model of particles, the effects of damping are most relevant at around or below the nHz frequencies. Our results showcase the importance of including the damping of primordial fluctuations in future analysis of induced gravitational waves. We provide detailed analytical formulas and approximations for the kernel of induced gravitational waves. Lastly, we discuss possible implications of damping in alleviating the gauge issue of induced gravitational waves and in suppressing the so-called poltergeist mechanism.
\end{abstract}

\maketitle

\section{Introduction}
According to the standard model of cosmology, the very early universe was dominated by a hot plasma of relativistic particles -- at leading order described by a homogeneous and isotropic perfect fluid. Cosmic Microwave Background (CMB) observations \cite{Planck:2018vyg} tell us that the radiation fluid had primordial fluctuations. On the largest scales, with wavelength $\lambda>{\rm Mpc}$, primordial fluctuations are Gaussian with an almost scale-invariant power spectrum. The leading explanation for the origin of primordial fluctuations is cosmic inflation (see Refs.~\cite{Brout:1977ix,Starobinsky:1979ty,Guth:1980zm,Sato:1981qmu} for the earliest works), which has been constrained to unprecedented precision at the largest angular scales \cite{Planck:2018jri}. 

On smaller scales though (i.e., corresponding to wavenumber $k\gtrsim 1\,{\rm Mpc}^{-1}$), current CMB and large-scale structure observations run out of steam because small-scale fluctuations are exponentially damped by dissipative effects \cite{Silk:1967kq, Weinberg:1971mx} (see also Refs.~\cite{Hu:1995em,dodelson2021modern}) and non-linearities scramble the primordial information. Constraints, therefore, weaken significantly or become much more uncertain. Robust constraints on the amplitude of scalar perturbations at $10\,{\rm Mpc}^{-1}\lesssim k\lesssim 10^4\,{\rm Mpc}^{-1}$ can be obtained using the {\it COBE/FIRAS} limits on $\mu$- and $y$-type CMB spectral distortions \citep{Chluba2012, Chluba2012Inflaton,Cyr:2023pgw}. At even smaller scales, various limits from Pulsar Timing and interferometric measurement can be derived (e.g., see Fig.~3 of \cite{Cyr:2023pgw}), currently still leaving a lot of room for small-scale perturbations with an amplitude that greatly exceeds that of the simplest slow-roll inflation models.

In contrast, with Gravitational Wave (GW) observations, we may be able to probe the primordial spectrum of fluctuations on the smallest scales \cite{Inomata:2018epa,Byrnes:2018txb}. This is because, in a perturbed (homogeneous and isotropic) universe, the evolution of density fluctuations generates anisotropic stresses, which then source GWs \cite{Tomita,Matarrese:1992rp,Matarrese:1993zf,Matarrese:1997ay} that travel to us almost unimpeded. The resulting GWs are often referred to as induced GWs \cite{Ananda:2006af,Baumann:2007zm} and are a crucial counterpart of the Primordial Black Hole (PBH) scenario \cite{Saito:2009jt,Bugaev:2009zh}. See, e.g., Refs.~\cite{Yuan:2021qgz,Domenech:2021ztg} for recent reviews and Ref.~\cite{Domenech:2024kmh} for a more pedagogical approach. Notably, Pulsar Timing Array (PTA) collaborations reported evidence of a nHz GW background \cite{EPTA:2023fyk,EPTA:2023sfo,EPTA:2023xxk,Zic:2023gta,Reardon:2023gzh,Reardon:2023zen,NANOGrav:2023hde,NANOGrav:2023gor,Xu:2023wog,InternationalPulsarTimingArray:2023mzf}, a possible interpretation of which are induced GWs \cite{NANOGrav:2023hvm,Franciolini:2023pbf,Figueroa:2023zhu,Inomata:2023zup,Liu:2023ymk,Ellis:2023oxs,Liu:2023pau,Domenech:2024rks}. See also Ref.~\cite{LISACosmologyWorkingGroup:2025vdz} for a study on the capabilities of LISA in the context of induced GWs.

Induced GWs have been studied in great detail in extensions of the simplest cosmology, that is, for a perfect fluid in a radiation-dominated universe with Gaussian primordial fluctuations. For example, beyond the first works on the topic one can take into account the effect of primordial non-Gaussianities \cite{Unal:2018yaa,Cai:2018dig,Atal:2021jyo,Adshead:2021hnm,Abe:2022xur,Garcia-Saenz:2023zue}, different expansion histories \cite{Inomata:2019ivs,Inomata:2019zqy,Papanikolaou:2020qtd,Domenech:2020ssp,Domenech:2021wkk,Inomata:2020lmk,Fernandez:2023ddy,Pearce:2023kxp}, isocurvature initial conditions \cite{Domenech:2021and,Domenech:2023jve,Domenech:2024wao,Marriott-Best:2025sez}, and modifications of gravity \cite{Zhang:2024vfw,Feng:2024yic,Domenech:2024drm,Tzerefos:2024rgb,Kugarajh:2025rbt}, to name just a few. All the resulting induced GW spectra, for a peaked primordial spectrum,\footnote{Two exceptions that lead to a resonant peak even if the primordial spectrum is scale-invariant are an early phase of matter domination \cite{Inomata:2019ivs,Inomata:2019zqy,Fernandez:2023ddy} or modified gravity \cite{Domenech:2024drm}.} share two common features: $(i)$ a resonant peak \cite{Ananda:2006af,Saito:2009jt,Pi:2020otn} and $(ii)$ a logarithmic running \cite{Cai:2018dig,Yuan:2019wwo} of the low frequency universal $f^3$ tail in a radiation dominated universe  \cite{Cai:2019cdl}, although it should be noted, that the logarithmic running is absent for a general equation of state of the early universe \cite{Domenech:2019quo,Domenech:2020kqm}. These features are most pronounced in the extreme case of an infinitely sharp primordial spectrum, namely a Dirac delta. In that case, the resonant peak becomes a logarithmic divergence and the low-frequency tail scales as $f^2\ln^2f$. The divergence is integrable and, therefore, often deemed unproblematic (also because observations have a finite bandwidth which removes the integrable divergence \cite{Thrane:2013oya,Byrnes:2018txb,Iovino:2024tyg}). However, from the theoretical point of view, this highlights an {\it unphysical} feature of the treatment.

Both effects, the resonant (divergent) peak and the logarithmic running have the same origin: \textit{a secular growth due to a slow decay of the source term}. However, this is an artefact of our simplifying assumptions, namely the perfect fluid approximation that neglects higher-order corrections. In reality, the plasma of relativistic particles is an imperfect fluid, with heat conductivity and viscosity, due to a finite mean free path of light particles \cite{Silk:1967kq,Weinberg:1971mx}. More concretely, the angular and polarization dependence of scattering processes in a tightly coupled fluid sources the quadrupole and isotropizes fluctuations \cite{Hu:1995em}. This causes dissipation and exponential damping of density fluctuations, introducing a {\it finite} effective lifetime of the primordial perturbation. 

An analogy from atomic physics comes to mind when thinking about this problem. It is well-known that the line profile for atomic dipole transitions (e.g., 2p-1s for Lyman-$\alpha$ in hydrogen) is approximately given by a Lorentzian. One crucial ingredient for removing the pole at the resonance is the finite lifetime of the 2p-state, which introduces an imaginary pole displacement due to damping. For GWs, a similar effect occurs when we consider damping due to viscosity as the main source of dissipation \cite{Jeong:2014gna}, though other dissipative channels like second-order density fluctuations and the GWs themselves are expected to contribute at higher order to the effective lifetime of the perturbation. From this picture one naturally expects additional GW ``line''-broadening effects, e.g., from bulk velocities, however, these are considered elsewhere.

In this paper, we show that dissipative effects regularize the induced GW spectrum. Namely, the amplitude of the resonant peak becomes finite and the logarithmic running stops below some break frequency. We provide an analytical form of the Kernel and accurate analytical approximations for computing the GW signal. We also find that dissipation at temperatures below the ElectroWeak (EW) phase transition has an important impact on the induced GW spectrum.  We note that the induced GW Kernel derived in this work is, in fact, a more realistic description of the induced GW generation in the early universe. Moreover, our formalism can also be used to probe the presence of very weakly interacting particles in the very early universe. 

It should be noted that, although Ref.~\cite{Yu:2024xmz} numerically investigated the effects of damping in the induced GW spectrum for a broad log-normal peak as a probe of new physics, we present an extensive and more general analytical study. We uncover the disappearance of the logarithmic running of the low-frequency tail, and the regularization of the resonant peak and show that damping in the standard model of particles could also be important around or below the nHz frequencies.

This paper is organized as follows. In Sec.~\ref{sec:induceddamping} we present the general derivation of the induced GW kernel in the presence of damping. We also provide the leading order contribution in a compact form. In Sec.~\ref{sec:powerlaw}, we focus on the case of a damping scale, which is a power-law in time, which is a good description for certain periods in the early universe. There, we derive analytical expressions for the general kernel. In Sec.~\ref{sec:spectrum}, we study the induced GW spectrum from peaked primordial spectra, focusing on the Dirac delta spectrum. We show the regularization of the resonant peak and the fading of the logarithmic running in the low-frequency tail. Lastly, we conclude our work in Sec.~\ref{sec:conclusions} and discuss future prospects. Many details on the derivation and approximations can be found in the Appendices.

\section{The Induced GW kernel in the presence of damping \label{sec:induceddamping}}
In the presence of dissipation, density fluctuations damp once they enter the Hubble radius. And so does the curvature perturbation. In the conformal Newton gauge\footnote{We follow the conventions for cosmological perturbation theory of Ref.~\cite{Domenech:2021ztg}. Namely, we take a perturbed FLRW metric in the Newton gauge given by
\begin{align}
ds^2=a^2(\eta)[-(1+2\Psi)d\eta^2+((1+2\Phi)\delta_{ij}+h_{ij})\,dx^idx^j]\,.
\end{align}
} (see Refs.~\cite{Kodama:1984ziu,Mukhanov:1990me,Bassett:2005xm, Baumann:2009ds} for reviews on cosmological perturbation theory) and in the very early, radiation-dominated universe, the Fourier $k$-mode of the Newtonian potential $\Phi$ in the presence of damping reads \cite{Jeong:2014gna,dodelson2021modern}
\begin{align}\label{eq:Phiktau}
\Phi(k,\tau)\approx\Phi_{\rm rad}(k,\tau)e^{-k^2/k_D^2(\tau)}\,,
\end{align}
where $\tau$ is conformal time,  $\Phi_{\rm rad}(k,\tau)$ is the solution without dissipation in the perfect fluid radiation-dominated universe and $k_D(\tau)$ is the damping scale. The damping factor appears since $\Phi$ is ultimately linked to the perturbations in the relativistic fluid (e.g., photons and neutrinos), which damp through dissipative effects. For super-Hubble initial conditions, $\Phi_{\rm rad}(k,\tau)$ is given by
\begin{align}
\Phi_{\rm rad}(k,\tau)=2\zeta_{{\rm prim},\mathbf{k}}\frac{j_1(c_sk\tau)}{c_sk\tau}\,,
\end{align}
where $c_s^2=1/3$ and $j_1(x)$ is the spherical Bessel function of order $1$. We also assumed that initial conditions are set by inflation on super-Hubble scales. As customary in this case, the initial conditions are given in terms of the primordial curvature fluctuations on uniform density slices $\zeta_{{\rm prim},\mathbf{k}}$. We then used that $\Phi=2/3\zeta$ on super-Hubble scales, i.e., at the initial time for all modes of interest (see, e.g., Ref.~\cite{Mukhanov:1990me}). Here, for simplicity, we take the primordial curvature fluctuations to be random Gaussian with dimensionless power-spectrum ${\cal P}_\zeta(k)$.\footnote{In our notation, the power-spectrum is defined by
\begin{align}
\langle\zeta_{{\rm prim},\mathbf{k}}\zeta_{{\rm prim},\mathbf{k}'}\rangle=\frac{2\pi^2}{k^3}{\cal P}_\zeta(k)\times(2\pi)^3\delta^{(3)}(\mathbf{k}+\mathbf{k}')
\end{align}}

The damping scale $k_D(\tau)$ in Eq.~\eqref{eq:Phiktau} is estimated via \cite{Jeong:2014gna}
\begin{align}\label{eq:k_D}
k_D^{-2}(\tau)=\int_0^\tau d\tilde\tau\,\Gamma_D(\tilde\tau)\approx \int_0^\tau d\tilde\tau\,\frac{2\eta(\tilde \tau)}{3a(\rho+P)}\,,
\end{align}
where $\Gamma_D(\tau)$ is the damping rate and $\rho$ and $P$ are the total energy density and pressure in the universe, respectively. Since the universe is dominated by radiation, we have that $P=\rho/3$. In the last step of Eq.~\eqref{eq:k_D}, we used that, in the very early universe, the damping rate is mainly dominated by the shear viscosity $\eta(\tau)$ \cite{ Weinberg:1971mx}. This is because at early times the fluid remains relativistic, tightly coupled and thermal, such that heat conduction and bulk viscosity remain suppressed \citep{Weinberg:1971mx,
Weinberg:2008}.  

Considering photons, neutrinos and possibly beyond standard model weakly interacting particles $X$, the shear viscosity is given by \cite{Jeong:2014gna,Yu:2024xmz}
\begin{align}
\eta(\tau)=\frac{16}{45}\rho_\gamma t_\gamma+\frac{4}{15}\sum_{j=X,\nu}\rho_jt_j\Theta(\tau_{j,\rm dec}-\tau)\,,
\end{align}
where $\rho_i$ and $t_i$, with $i=\{\gamma,\nu,X\}$, are respectively the energy densities and mean free scattering time of photons, neutrinos and $X$-particles. $\tau_{j,\rm dec}$ denotes the decoupling time of $\nu$ and $X$. The mean free scattering times are computed via
\begin{align}
t_\gamma = (n_{e^\pm}\sigma_{\rm KN})^{-1}\,,\, t_\nu=(n_{\nu}\sigma_\nu)^{-1} \,{\rm and} \,,\, t_X=(n_{X}\sigma_X)^{-1}\,,
\end{align}
where the $\sigma_j$’s are the thermally averaged cross-sections ($\sigma_{\rm KN}$ is the Klein-Nishina cross section, $\sigma_\nu$ is the weak interaction cross-section and $\sigma_X$ is arbitrary). $n_j$’s are the number densities of $e^{\pm}$, $\nu$ and $X$. In what follows, we will be concerned about temperatures above ${\rm MeV}$. So, we can neglect the contribution from the photons' shear viscosity, as the weak interaction has the largest mean free path. In addition to damping scalar perturbations, free-streaming neutrinos also directly lead to the damping of GWs as they propagate through the universe \citep{Weinberg2004, Dicus2005}, imprinting features in the primordial GW spectrum \citep{Watanabe2006, Hook:2020phx, Kite2022GW,Franciolini:2023wjm} (see Refs.~\cite{Baumann:2007zm,
Abe:2020sqb} for works specific to the induced GWs); however, we do not consider these aspects in more detail here.

Before proceeding, it is instructive to provide numerical estimates for $k_D(\tau)$. For clarity, we rewrite Eq.~\eqref{eq:k_D} as\footnote{We used that for ultra-relativistic particles in thermodynamical equilibrium at temperature $T$ one has
\begin{align}
n_j=c_n\frac{\zeta(3)}{\pi^2}g_jT^3\quad{\rm and}\quad \rho_j=c_\rho\frac{\pi^2}{30}g_jT^4\,,
\end{align}
where for fermions $c_n=3/4$ and $c_\rho=7/8$, while for bosons $c_n=c_\rho=1$. We also used that $3H^2M_{\rm pl}^2=\rho_{\rm tot}(T)$ and that $g_s(T)a^3T^3={\rm constant}$ to relate $a$ with $T$.}
\begin{align}\label{eq:k_D2}
k_D^{-2}(T)\approx \,5.2\times &10^{-35}{\rm Mpc}^2\int_0^{\ln a} d\ln \tilde a \,\Theta(T-{\rm MeV})\nonumber\\&\times\left(\frac{g_s(T)}{106.75}\right)^{2/3}\left(\frac{g_\rho(T)}{106.75}\right)^{-3/2}\sum_j\frac{c_{\rho j}}{c_{n j}}\left(\frac{\sigma_{j,\rm EW}}{G_F^2T_{\rm EW}^2}\right)^{-1}\left(\frac{T}{T_{\rm EW}}\right)^{-3+m_j(T)} \,,
\end{align}
where $a$ is the scale factor and we chose the scale of EW symmetry breaking as a reference, here at $T_{\rm EW}=100\,{\rm GeV}$, $G_F\sim 10^{-5}\,{\rm GeV}^{-2}$ is the Fermi constant and we approximated $\sigma_j$ as power-laws, namely
\begin{align}\label{eq:sigmajT}
\sigma_j(T)=\sigma_{j,\rm EW}\,(T/T_{\rm EW})^{m_j(T)}\,.
\end{align}
The Heaviside theta function in Eq.~\eqref{eq:k_D2} takes into account that neutrinos decouple at $T\sim 1\,{\rm MeV}$. In our parametrization \eqref{eq:sigmajT}, we assume that $m_j(T)$ is a constant. The $T$ dependence in $m_j(T)$ is to denote that depending on the interactions, the cross-section may have different temperature dependence. For example, for gauge interactions $m_j=2$, for massive mediators $m_j=-2$ (see below). Whenever the value of $m_j(T)$ changes, we require continuity of $\sigma_j(T)$. 

For the weak interaction, we have that $\sigma_{\nu}(T<T_{\rm EW})\sim G_F^2T^2$, where $G_F\sim \alpha/M_W^2$ (with $M_W$ being the W-boson mass), and $\sigma_{\nu}(T>T_{\rm EW})\sim \sigma_{\nu}(T_{\rm EW})$ since we have now a cross-section dominated by gauge interactions, basically leading to $G_F\sim T^{-2}$. Thus, $m_\nu=2$ for $T<T_{\rm EW}$ and $m_\nu=-2$ for $T>T_{\rm EW}$. These regimes respectively correspond to massive or massless W-bosons, assuming an instantaneous cross-over transition. Since the integrand is dominated by the late time contribution, and we have that $a\propto T^{-1}$, a good approximation to $k_D$ is given by 
\begin{align}\label{eq:k_Dappr}
k_D(T)\approx \,1.4\times &10^{17}{\rm Mpc}^{-1}\nonumber\\&\times\sqrt{3-m_j(T)}\sqrt{\frac{\sigma_{j,\rm EW}}{G_F^2T_{\rm EW}^2}}\left(\frac{T}{T_{\rm EW}}\right)^{\frac{3-m_j}{2}}\left(\frac{g_s(T)}{106.75}\right)^{-1/3}\left(\frac{g_\rho(T)}{106.75}\right)^{3/4}  \,,
\end{align}
where we assumed that only one particle species, say only $j=\nu$ or $j=X$, dominates the shear viscosity in specific temperature ranges and we neglected numerical factors due to fermions/bosons. If neutrinos dominate the viscosity we have that $k_D(T_{\rm EW})\approx \,1.4\times 10^{17}{\rm Mpc}^{-1}$. For $T<T_{\rm EW}$ we have $k_D(T)\propto T^{5/2}$ and for $T>T_{\rm EW}$ then $k_D(T)\propto T^{1/2}$. If the $X$ particles dominate the shear viscosity then $m_X(T)$ may have different values and change at different temperatures. The estimate \eqref{eq:k_Dappr}, though, is still valid when evaluated at a fixed temperature.

\begin{figure}
\includegraphics[width=0.48\columnwidth]{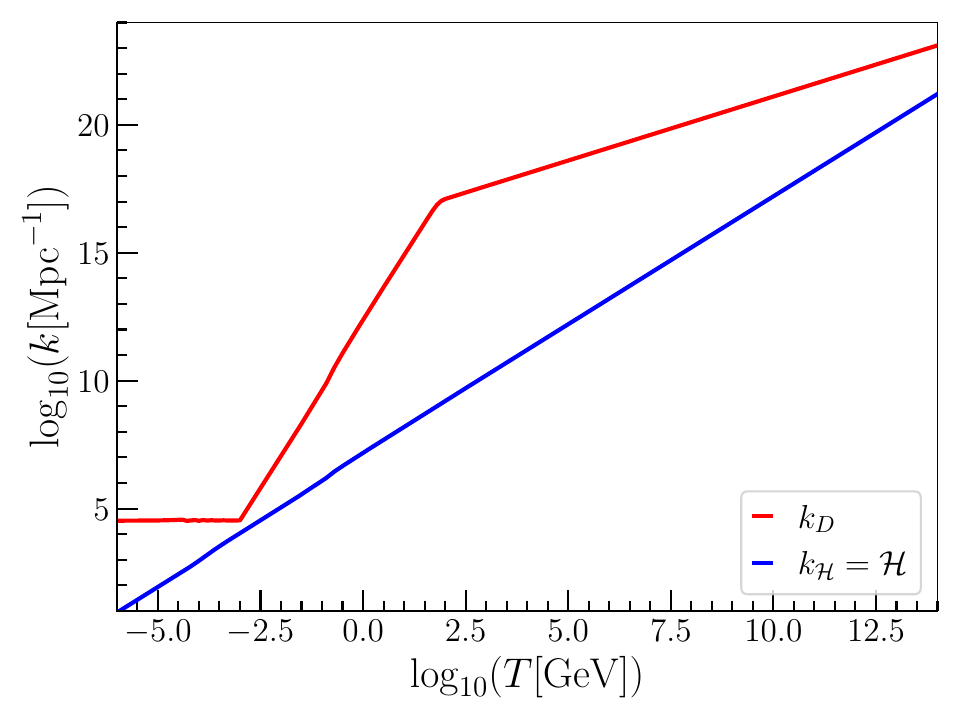}
\includegraphics[width=0.51\columnwidth]{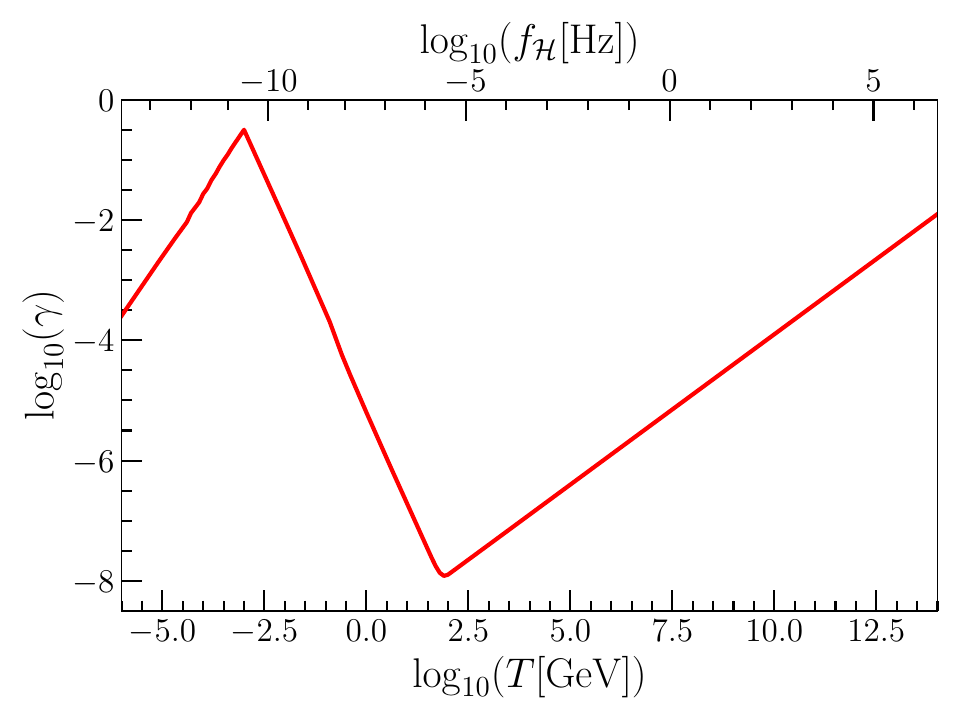}
\caption{On the left panel, we show the comoving damping scale $k_{D}$ \eqref{eq:k_D2} and the comoving Hubble scale ${\cal H}$ \eqref{eq:H(T)} as a function of temperature assuming the standard models of particles. After the EW phase transition, the damping scale decays with $T^{-5/2}$, approaching the Hubble scale. Then it becomes a constant after neutrinos decouple until dissipations by photons take over the damping at $T\sim {\rm keV}$. On the right panel, we show the ratio of scales $\gamma$ \eqref{eq:gammaTex} as a function of $T$. In the upper $x$-axis, we give the frequency corresponding to the Hubble radius at that temperature, namely $f_{\cal H}={\cal H}/(2\pi)$. See how $\gamma$ is bounded by $10^{-8}\lesssim \gamma \lesssim 0.3$. Note that, though the functional form of $\gamma$ will change in the presence of $X$, the general bounds will not.\label{fig:kdgamma}}
\end{figure}

One relevant quantity for our study is the ratio of the damping scale and the Hubble scale at a given time. 
For reference, the comoving Hubble scale ${\cal H}=a'/a$, where $'=d/d\tau$, is given by
\begin{align}\label{eq:H(T)}
{\cal H}(T)\approx 1.6 \times 10^9\,{\rm Mpc}^{-1}\left(\frac{T}{T_{\rm EW}}\right)\left(\frac{g_s(T)}{106.75}\right)^{-1/3}\left(\frac{g_\rho(T)}{106.75}\right)^{1/2}\,.
\end{align} 
Then, from Eqs.~\eqref{eq:k_Dappr} and \eqref{eq:H(T)} we have that
\begin{align}\label{eq:gammaTex}
\gamma(T)\equiv\frac{{\cal H}(T)}{k_D(T)} \approx\frac{1.2\times 10^{-8}}{\sqrt{3-m_j(T)}}\left(\frac{\sigma_{j,\rm EW}}{G_F^2T_{\rm EW}^2}\right)^{-1/2}\left(\frac{T}{T_{\rm EW}}\right)^{\frac{m_j-1}{2}}\left(\frac{g_\rho(T)}{106.75}\right)^{-1/4}\,.
\end{align} 
Note that since $m_\nu=2$ for $T<T_{\rm EW}$, so $\gamma\propto T^{1/2}$, and $m_\nu=-2$ for $T>T_{\rm EW}$, so $\gamma\propto T^{-3/2}$, we have that  $\gamma(T_{\rm EW})\sim 10^{-8}$ is a lower bound. Thus, for the standard model of particles, we have that, in general, $\gamma(T)\gtrsim 10^{-8}$. In the presence of beyond standard model particles $X$, the lower bound of $\gamma(T)$ could be larger if $\sigma_{X,\rm EW}<G_F^2T_{\rm EW}^2$. As we will see, $\gamma(T)$ will be related to the scale below which the induced GWs lose the logarithmic running in the low-frequency tail. Thus, the larger the $\gamma$ the sooner the effects appear. In Fig.~\ref{fig:kdgamma}, we show the temperature dependence of the damping scale as well as $\gamma$, assuming the standard model of particles. Note how the strongest damping happens at temperatures corresponding to around nHz frequencies.


\subsection{The induced GW spectrum}
To estimate the spectrum of induced GWs in the presence of damping we follow the procedure introduced in Ref.~\cite{Kohri:2018awv,Espinosa:2018eve,Domenech:2019quo,Domenech:2020kqm} (see also Refs.~\cite{Yuan:2021qgz,Domenech:2021ztg} for reviews). For Gaussian primordial fluctuations, the spectrum of induced GWs is given by \cite{Domenech:2021ztg} 
\begin{align}\label{eq:Phgaussian}
{\cal P}_h(k,\tau)=8\int_0^\infty\id v\int_{|1-v|}^{1+v} \id u\left[\frac{4v^2-(1-u^2+v^2)^2}{4uv}\right]^2 I^2(k,\tau,v,u)\,{{\cal P}_{\zeta}(ku)}\,{{\cal P}_{\zeta}(kv)}\,,
\end{align}
where $v$ and $u$ are related to the internal momenta of the curvature fluctuations. The kernel, or transfer function, $I(k,\tau,u,v)$ is defined by
\begin{align}\label{eq:Idefinition}
I(k,\tau,u,v)=k\int_{0}^{\infty}  d\tilde \tau\,G_h(k,\tau, \tilde\tau)\,f(\tilde \tau,k,u,v)\,,
\end{align}
where $G_h(k,\tau,\tilde \tau)$ is the Green’s function for the tensor modes\footnote{In the radiation dominated universe it is given by 
\begin{align}
G_h(k,\tau,\tilde \tau)=(k\tilde \tau)^2\left[y_0(k\tau)j_0(k\tilde \tau)-j_0(k\tau)y_0(k\tilde \tau)\right]\,,
\end{align}
where $j_0(x)$ and $y_0(x)$ are the spherical Bessel functions of order 0 of the first and second kind respectively.} 
and $f(k,\tau,u,v)$ is the second order source to the equation of motion of tensor modes $h_{ij}$. For the case at hand, it is given by
\begin{align}\label{eq:f}
f(k,\tau, u,v)=&T_\Phi(uk,\tau)T_\Phi(vk,\tau)+\frac{1}{2}\left(T_\Phi(uk,\tau)+\frac{T'_\Phi(uk,\tau)}{\cal H}\right)\left(T_\Phi(vk,\tau)+\frac{T'_\Phi(vk,\tau)}{\cal H}\right)\,,
\end{align}
where we introduced $T_\Phi(k,\tau)$ as the transfer function for $\Phi(k,\tau)$ in Eq.~\eqref{eq:Phiktau}, namely
\begin{align}
T_\Phi(k,\tau)=2\,\frac{j_1(c_sk\tau)}{c_sk\tau}\,e^{-\kappa_{D}^2F[\tau/\tau_*]}\,,
\end{align}
which here includes the damping terms.
In the transfer function, we have for convenience introduced an arbitrary reference time $\tau_*$ and defined the following dimensionless quantities,
\begin{align}
\kappa_{D}\equiv \frac{k}{k_D(\tau_*)}\quad {\rm and}\quad F[\tau/\tau_*]\equiv k^2_D(\tau_*)/k^2_D(\tau)\,.
\end{align}
Note that in Eq.~\eqref{eq:Idefinition} we take the $\tau\to\infty$ limit of the integral since we are interested in modes which entered the Hubble radius much before the radiation-matter equality. Lastly, with the power spectrum given by Eq.~\eqref{eq:Phgaussian}, the induced GW spectral density is
\begin{equation}\label{eq:spectraldensity}
\Omega_{\rm GW}(k)=\frac{1}{3 M_{\rm pl}^2 H^2}\frac{\id \rho_{\rm GW}}{\id\ln k}=\frac{k^2}{12a^2H^2}\,\overline{{\cal P}_h(k,\tau)}\,,
\end{equation}
where the overline denotes the oscillation average today. This basically replaces $\cos^2x$ and $\sin^2x$ terms by $1/2$ and kills the cross terms in the square of Eq.~\eqref{eq:Idefinition}. For later use, the induced GW spectral density today is given by
\begin{align}\label{eq:spectraldensitytoday}
\Omega_{\rm GW,0}h^2=1.62\times 10^{-5}\left(\frac{\Omega_{r,0}h^2}{4.18\times 10^{-5}}\right)\left(\frac{g_*(T_{\rm c})}{106.75}\right)\left(\frac{g_{*s}(T_{\rm c})}{106.75}\right)^{-4/3}\Omega_{\rm GW,c}(k)\,,
\end{align}
where the subscript ``c'' denotes evaluation at a time when the GWs are sufficiently deep inside the Hubble radius such that $\Omega_{\rm GW,c}$ for a given $k$-mode is effectively constant \cite{Inomata:2016rbd}. For simplicity, we will take this to be the time of Hubble entry, though to be more precise it should be a couple of e-folds after that \cite{Pritchard:2004qp,Smith:2006nka}. We will also drop the subscript  ``c'' from now on, unless strictly necessary.

\subsection{The general induced GW kernel}

We proceed to simplify the form of the kernel \eqref{eq:Idefinition}. For convenience, we split the kernel into
\begin{align}\label{eq:generalIkernel}
I(k,\tau,u,v)=I_{y}(k,u,v)j_0(k\tau)+I_{j}(k,u,v) y_0(k\tau)
\end{align}
where
\begin{align}\label{eq:Ijydef}
I_{j/y}(k,u,v)=\int_{0}^{\infty} d\tilde x\, \tilde x^2\,\left\{
\begin{aligned}
j_0(\tilde x)\\y_0(\tilde x)
\end{aligned}
\right\}
f(k,\tilde x,u,v)\quad {\rm with}\quad x=k\tau\,;\,\tilde x=k\tilde\tau.
\end{align}
We further split $I_j$ and $I_y$ into terms containing no derivatives of $F[\tau/\tau_*]$, one derivative and one derivative squared, respectively
\begin{align}\label{eq:IjIygamma}
I_j= I^{(0)}_j+\gamma_*^2I^{(1)}_j+\gamma_*^4I^{(2)}_j\quad ,\quad I_y= I^{(0)}_y+\gamma_*^2I^{(1)}_y+\gamma_*^4I^{(2)}_y\,,
\end{align}
where we defined
\begin{align}\label{eq:gamma}
\gamma_*=\gamma(T_*)=1/(k_{D*}\tau_*)=k_*/k_{D*}\ll 1\,,
\end{align}
and in the last two steps, we used that in the radiation-dominated universe ${\cal H}=1/\tau$, and so the mode corresponding to the Hubble radius at the pivot time is $k_*={\cal H}_*=1/\tau_*$. Recall that $\gamma(T)$ is defined in Eq.~\eqref{eq:gamma}. Note that from now on we will suppress the arguments in $I_j(k,u,v)$ and $I_y(k,u,v)$ for simplicity. The exact form of $I^{(m)}_j$ and $I^{(m)}_y$, with $m=\{0,1,2\}$, are given in App.~\ref{app:detailskernel}. The most important point is that the terms proportional to $\gamma_*$ in Eq.~\eqref{eq:IjIygamma} are in general suppressed, as we will show later in Sec.~\ref{sec:corrections}.

After several integrations by parts, and at leading order in $\gamma_*$, the induced GW kernels $I_j$ and $I_y$ are approximately given by
\begin{align}\label{eq:Ijapprox}
I_j\approx&\,{\cal I}_j^{(0)}=-\frac{1-c_s^2 \left(u^2+v^2\right)}{2 c_s^4 u^2 v^2}\left[1
   -\frac{1-c_s^2 \left(u^2+v^2\right)}{4 c_s^2 u v}\left({\rm cei}[y_1]+{\rm cei}[y_2]-{\rm cei}[y_3]-{\rm cei}[y_4]\right)\right]\,,
\end{align}
and
\begin{align}\label{eq:Iyapprox}
I_y\approx &\,{\cal I}_y^{(0)}=
   \frac{\left[1-c_s^2 \left(u^2+v^2\right)\right]^2}{8 c_s^6 u^3 v^3}\left({\rm Sei}[y_1]+{\rm Sei}[y_2]-{\rm Sei}[y_3]-{\rm Sei}[y_4]\right)\,,
\end{align}
where, for compactness, we have introduced
\begin{align}\label{eq:yjdefinition}
y_1= 1-c_s(u-v)\,,\, y_2= 1+c_s (u-v)\,,\, y_3=1-c_s (u+v)\,,\, y_4=1+c_s(u+v)\,.
\end{align}
Note that since $|u-v|\leq1$ and $1\leq u+v<\infty$, which results from  $|1-v|<u<1+v$, then it follows that
\begin{align}
1-c_s\leq y_1,y_2\leq 1+c_s\quad,\quad y_4\geq1+c_s \quad{\rm and}\quad -\infty<y_3<1-c_s\,.
\end{align}
We have also defined for convenience
\begin{align}\label{eq:cei01}
{\rm cei}[y_i]=\int_{0}^\infty \frac{dx}{x} e^{-{(u^2+v^2)\kappa_{D}^2}F[x/x_*]} \left[1-\cos (y_ix)\right]
\,,
\end{align}
and
\begin{align}\label{eq:Sei01}
{\rm Sei}[y_i]=\int_{0}^\infty \frac{dx}{x} e^{-{(u^2+v^2)\kappa_{D}^2}F[x/x_*]} \sin (y_ix)\,.
\end{align}
The notation ``${\rm cei}$'' and ``${\rm Sei}$'' stands for generalized cosine and sine exponential integral, respectively. Note that in the expressions above $x_*=k\tau_*$. Also, note that in the definition of ${\rm cei}[y_i]$ \eqref{eq:cei01} we added a factor $1$ so that the integral converges, as it removes the divergence at $x=0$. This is usual practice with the cosine integrals; see, e.g., Ref.~\cite{NIST:DLMF} Eq.~6.2.12. In the kernel \eqref{eq:Ijapprox}, the factors $1$ in ${\rm cei}[y_i]$ \eqref{eq:cei01} cancel each other, but the individual integrals are better behaved, numerically and analytically. Nevertheless, the form of Eqs.~\eqref{eq:Ijapprox} and \eqref{eq:Iyapprox} is more practical for analytical investigation.\footnote{
Although the full numerical investigation is out of the scope of this work, we note that the sum of integrals in Eq.~\eqref{eq:Ijapprox} combines to
\begin{align}
{\rm cei}[y_1]+{\rm cei}[y_2]-{\rm cei}[y_3]-{\rm cei}[y_4]=-4\int_0^\infty\frac{dx}{x} e^{-{(u^2+v^2)\kappa_{D}^2}F[x/x_*]}\cos x\sin(c_sux)\sin(c_svx)\,.
\end{align}
In Eq.~\eqref{eq:Ijapprox} one obtains the same integral but for a replacement of $\cos x\to \sin x$ and $-4\to 4$. This approach might be more practical for numerical implementation as it involves a single integral.} We give the exact expressions for $I_j$ and $I_y$ after integration by parts as well as more details on the trigonometric exponential integrals, as in Eqs.~\eqref{eq:cei01} and \eqref{eq:Sei01}, in App.~\ref{app:detailskernel}. 
Unfortunately, Eqs.~\eqref{eq:Ijapprox} and \eqref{eq:Iyapprox} are only tractable numerically in the general case. To obtain an analytical understanding, we focus on the case where $F\propto \tau^\alpha$ in the next section, which is well motivated by particle physics models, see Eq.~\eqref{eq:k_D2}. As we shall see, in some cases the integrals can be done explicitly.

Before that, as a sanity check, we confirm that when $F[x/x_*]=0$ (or, equivalently, $k_{D}\to\infty$ and $\gamma_*\to 0$) we recover the expressions of Ref.~\cite{Domenech:2021ztg}, which coincides with Kohri and Terada \cite{Kohri:2018awv} after taking into account differences in convention. In particular, we have that\footnote{For $F[x/x_*]=0$, the integral ${\rm cei}[y_i]$ does not converge but the integral of the difference of ${\rm cei}[y_i]$ does. This is clear by setting a lower cut-off to the integral and taking the limit to zero after integration.}
\begin{align}
{\rm cei}[y_i;k_D\to\infty]-{\rm cei}[y_j;k_D\to\infty]=\ln\left|y_i/y_j\right|
\quad{\rm and}\quad
{\rm Sei}[y_i;k_D\to\infty]={\rm sign}[y_i]\frac{\pi}{2}\,.
\end{align}
Thus, after some algebraic manipulations, we find
\begin{align}\label{eq:Ijnodamp}
I_j(k_{D}\to\infty)=-\frac{\left[1-c_s^2 \left(u^2+v^2\right)\right]}{8 c_s^6 u^3 v^3} \left\{4 c_s^2 u v-
   \left[1-c_s^2 \left(u^2+v^2\right)\right]\ln \left|\frac{1-c_s^2 (u-v)^2}{1-c_s^2
   (u+v)^2}\right|\right\}\,,
\end{align}
and
\begin{align}\label{eq:Iynodamp}
I_y(k_{D}\to\infty)=\frac{\pi  \left[1-c_s^2 \left(u^2+v^2\right)\right]^2 }{8 c_s^6 u^3 v^3}\,\Theta(c_s
   (u+v)-1)\,,
\end{align}
which coincide with Ref.~\cite{Domenech:2021ztg}.

\section{The case of power-law damping scale\label{sec:powerlaw}}
In the history of the very early universe there were periods in which only one interaction dominates the damping scale, at least this is the case if we assume the standard model of particles (see Eq.~\eqref{eq:k_D2} and the discussion around it). During that time, the damping scale is very well approximated by a power-law. Let us then fix
\begin{align}\label{eq:F}
F[x/x_*]=(x/\kappa)^\alpha\quad{\rm with}\quad \kappa\equiv k/k_*\,,
\end{align}
where we used that $x_*=k\tau_*=k/k_*=\kappa$. We also consider $\alpha\geq0$.  Namely, we assume that the damping scale, that goes as $k_D\propto F^{-1/2}$, is a decreasing function of time. This follows from the definition \eqref{eq:k_D} since $\Gamma_D>0$. We note that the case of $\alpha=0$ is rather trivial, and from particle physics interactions, we find that $\alpha\geq1$.\footnote{Note that for $\alpha>1$ the exponential damping is valid as long as $k_{D}(\tau)> {\cal H}$. This means that the power-law approximation is only valid until for
\begin{align}
x<\kappa\gamma_*^{-1/(\alpha-1)}\,.
\end{align}
Since $\gamma_*\ll1$ this is not a problem per se as long as the primordial power-spectrum has a cut-off at high-momenta satisfying $k_{\rm cut}/k_*<\gamma_*^{-1/(\alpha-1)}$. We could also consider the precise evolution of $k_D(\tau)$ \eqref{eq:k_D2} to avoid this issue. For $\alpha<1$, we have the opposite situation, meaning that if the condition $k_{D}(\tau)> {\cal H}$ is initially satisfied, then it is always satisfied.}
Thus, we will mainly focus on $\alpha\geq1$ from now on. We leave a detailed study for $\alpha<1$ for future work, although some of the formulas derived in this paper also apply.

For a fixed value of $\alpha$, we can integrate by parts the Kernel \eqref{eq:IjIygamma} until the highest inverse power of $x$ in the integrals \eqref{eq:Ijydef} is $x^{-1}$ for $\alpha\in{\mathbb{Z}}^+$ or $x^\delta$ where $0<\delta<1$ for $\alpha\notin{\mathbb{Z}}^+$. This means that we can further split the Kernel \eqref{eq:IjIygamma} after more integration by parts, as
\begin{align}\label{eq:calIjIy}
I_j={\cal I}^{(0)}_j+\sum_{m=1}^{m_{\rm max}} \gamma_*^{2m}{\cal I}^{(m)}_j\quad,\quad I_y={\cal I}^{(0)}_y+\sum_{m=1}^{m_{\rm max}} \gamma_*^{2m}{\cal I}^{(m)}_y\,,
\end{align}
where $m\in{\mathbb{Z}}^+$. For $\alpha\in {\mathbb{Z}}^+$ we have that $m_{\rm max}=4$ for $\alpha=1$, while $m_{\rm max}=2$ for $\alpha\geq 2$. For $\alpha\notin{\mathbb{Z}}^+$, $m_{\rm max}={\rm Max}\left\{2,{\rm Ceiling}[5/\alpha]\right\}$.\footnote{Before integration by parts we have roughly integrals of the type $\int dx\, \gamma^2 x^{\alpha-5} e^{-\gamma^2 x^\alpha}$ (see App.~\ref{app:detailskernel}). After the $m$-th integration by parts we obtain terms with $\int dx\, \gamma^{2m} x^{m\alpha-5} e^{-\gamma^2 x^\alpha}$. These integrals, or the combination of integrals with the same powers of $x$, converge at $x\to 0$ when $m\alpha-5\geq0$. For $\alpha\in{\mathbb{Z}}^+$, the maximum power one can obtain without involving logarithms is $m\alpha-5=-1$. }
Note that the terms ${\cal I}^{(0)}_j$ and ${\cal I}^{(0)}_y$ only depend on $\alpha$ through the exponential. The terms ${\cal I}^{(m)}_j$ and ${\cal I}^{(m)}_j$ instead have additional powers of $x^\alpha$ as they depend on derivatives of the exponential. Let us then study the general behavior of ${\cal I}^{(0)}_j$ and ${\cal I}^{(0)}_y$, which is roughly the same for all $\alpha$, and then turn our attention to ${\cal I}^{(m)}_j$ and ${\cal I}^{(m)}_j$. 

\subsection{Common terms and their asymptotics}
Although we wrote the expressions for ${\cal I}^{(0)}_j$ and ${\cal I}^{(0)}_y$ in Eqs.~\eqref{eq:Ijapprox} and \eqref{eq:Iyapprox}, let us write them explicitly here for consistency. We also find it more convenient, both numerically and analytically, to rewrite the integrals \eqref{eq:cei01} and \eqref{eq:Sei01} as
\begin{align}\label{eq:cei012}
{\rm cei}[\beta y_j,\alpha]=\int_{0}^\infty \frac{d z}{ z} e^{- z^\alpha} \left[1-\cos (\beta y_j  z)\right]\quad{\rm and}\quad{\rm Sei}[\beta y_j,\alpha]=\int_{0}^\infty \frac{d z}{ z} e^{- z^\alpha} \sin (\beta y_j  z)\,,
\end{align}
where we changed the integration variable to $x=\beta  z$ and we defined
\begin{align}\label{eq:betaj}
\beta =\frac{\kappa}{[\kappa_{D}^{2}(u^2+v^2)]^{1/\alpha}}=\frac{\kappa^{1-2/\alpha}}{[\gamma_*^2(u^2+v^2)]^{1/\alpha}}\,.
\end{align}
In the last step of Eq.~\eqref{eq:betaj} we used that $\kappa_{D}=\gamma_*\kappa$, which follows from Eq.~\eqref{eq:gamma}. With these variables, we have that
\begin{align}\label{eq:calIj0}
{\cal I}^{(0)}_j=&-\frac{1-c_s^2 \left(u^2+v^2\right)}{2 c_s^4 u^2 v^2}\left[1
   -\frac{1-c_s^2 \left(u^2+v^2\right)}{4 c_s^2 u v}\left({\rm cei}[\beta y_1]+{\rm cei}[\beta y_2]-{\rm cei}[\beta y_3]-{\rm cei}[\beta y_4]\right)\right]\,,
\end{align}
and
\begin{align}\label{eq:calIy0}
{\cal I}^{(0)}_y= &
   \frac{\left[1-c_s^2 \left(u^2+v^2\right)\right]^2}{8 c_s^6 u^3 v^3}\left({\rm Sei}[\beta y_1]+{\rm Sei}[\beta y_2]-{\rm Sei}[\beta y_3]-{\rm Sei}[\beta y_4]\right)\,.
\end{align}
Note that here, as opposed to Eqs.~\eqref{eq:Ijapprox} and \eqref{eq:Iyapprox}, we wrote explicitly that the “cei” and “Sei” integrals are a single function of $\beta y_j$, for a fixed value of $\alpha$. Namely, the damping only enters through a single parameter, $\beta$. We are now ready to study the asymptotic behaviour of the Kernel. 

\begin{figure}
\includegraphics[width=0.49\columnwidth]{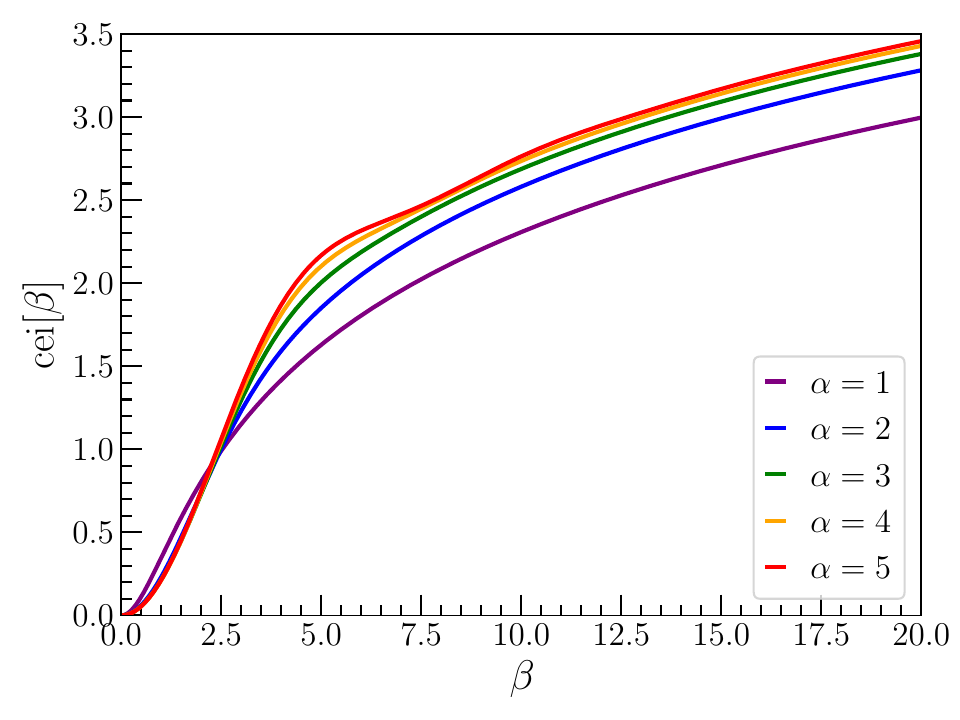}
\includegraphics[width=0.49\columnwidth]{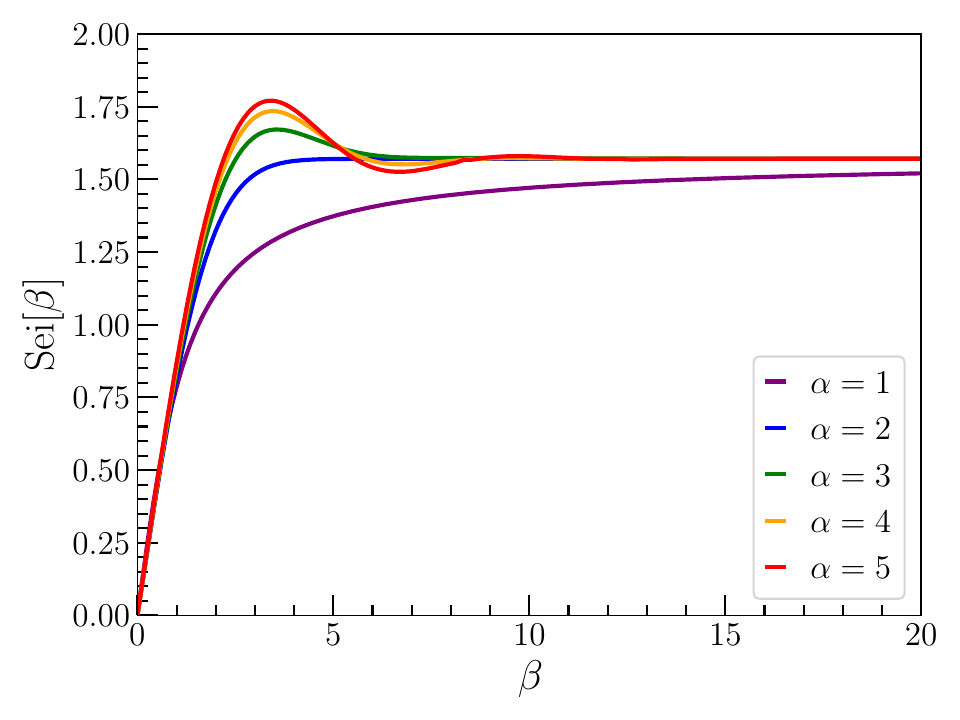}
\caption{We show the ${\rm cei}[\beta]$ (on the left) and ${\rm Sei}[\beta]$ (on the right) integrals given in Eq.~\eqref{eq:cei012} for various values of $\alpha$ as a function of $\beta$ (after absorbing $y_j$, namely $\beta y_j\to \beta$). In purple, blue, green, orange and red we show the cases $\alpha=1,2,3,4,5$, respectively. We only show the region $\beta>0$ as $\beta<0$ follows from symmetry. In particular, ${\rm cei}[\beta]$ is even and ${\rm Sei}[\beta]$ is odd. Note that, as $\alpha$ increases, the ${\rm cei}[\beta]$ and ${\rm Sei}[\beta]$ integrals resemble the shape of cosine and sine integrals, respectively. \label{fig:ceiandSei}}
\end{figure}

The two relevant limiting cases are $\beta y_j\ll1$ and $\beta y_j\gg1$. On one hand, $\beta y_j\ll1$ corresponds to either $y_j\ll1$, which is only possible when $c_s(u+v)\sim1$ at the resonant frequency, or a strong damping that is $\beta \ll1$. On the other hand, $\beta y_j\gg 1$ corresponds to undamped modes. We find that asymptotic behaviors of ${\rm cei}$ and ${\rm Sei}$, see App.~\ref{app:exponentialintegral}, are respectively given by
\begin{align}\label{eq:ceiasympt}
{\rm cei}[\beta y_j,\alpha]\approx\left\{
\begin{aligned}
&\Gamma\left[1+2{\alpha}^{-1}\right](\beta y_j)^2/4&\beta y_j\ll 1\\
&\ln|\beta y_j|+\left(1-\alpha^{-1}\right)\gamma_E&\beta y_j\gg 1
\end{aligned}
\right.\,,
\end{align}
where $\gamma_E\approx 0.577$ is Euler’s constant and
\begin{align}\label{eq:Seiasympt}
{\rm Sei}[\beta y_j,\alpha]\approx\left\{
\begin{aligned}
&\Gamma\left[1+{\alpha}^{-1}\right]\beta y_j&\beta y_j\ll 1\\
&\frac{\pi}{2}{\rm sign}[y_j]&\beta y_j\gg 1
\end{aligned}
\right.\,.
\end{align}
More accurate asymptotic expressions are given in App.~\ref{app:exponentialintegral}.
Thus, we see that in the limit $\beta y_j\gg1$ we recover the logarithmic terms in \eqref{eq:calIj0} and \eqref{eq:calIy0}, which are responsible for the logarithmic divergence at the resonant frequency and the logarithmic running of the induced GWs \cite{Cai:2018dig,Yuan:2019wwo}. However, when $\beta y_j\ll1$ the Kernels become suppressed by positive powers of $\beta y_j$. This has two important implications for peaked primordial spectra. First, there is no real divergence in the kernel at the resonant frequency and, second, that below some frequency, there is no logarithmic running. This was to be expected since both features stem from a secular growth integrated over infinitely long periods.\footnote{It is straightforward to check the previous statement from the full analytic expression of the kernel given, e.g., in Ref.~\cite{Kohri:2018awv} and then sending $\tau\to\infty$ in the integral \eqref{eq:Idefinition}.} Before studying the induced GW spectrum for a peaked primordial spectrum in Sec.~\ref{sec:spectrum}, let us analytically investigate the Kernel in the presence of damping more thoroughly.

\subsection{Exact and approximate integrals and fitting functions \label{ref:exactandapproximate}} 
In the previous section we have presented the functional form of the Kernel in the asymptotic limits in Eqs.~\eqref{eq:ceiasympt} and \eqref{eq:Seiasympt}. Nevertheless, when $\alpha=1$ and $\alpha=2$ there are exact analytical formulas. Let us show them here. First, for $\alpha=1$, we find that
\begin{align}\label{eq:seiandcei}
{\rm Sei}[z,\alpha=1]=\tan^{-1}[z]
\quad{\rm and}\quad
{\rm cei}[z,\alpha=1]=\frac{1}{2}\ln\left({1+z^2}\right)\,.
\end{align}
While for $\alpha=2$ we obtain\footnote{By matching the asymptotic behaviours we find an empirical fit, accurate to $1\%$, given by
\begin{align}
{\rm cei}[z,\alpha=2]={\rm IDi}[z]=\int^{z}_{0}dx\, {\rm Di}[x/2]\approx\frac{z^2}{4}e^{-z^2/12}+\frac{1}{2}\left(1 - e^{-z^2/14}\right)^3 \left(\gamma_E+2\ln\left|z\right|\right) \,.
\end{align}
}
\begin{align}\label{eq:IDidef}
{\rm Sei}[z,\alpha=2]=\frac{\pi}{2}{\rm Erf}\left[\frac{z}{2}\right]
\quad{\rm and}\quad
{\rm cei}[z,\alpha=2]=\int^{z}_{0}dx\, {\rm Di}[x/2]\equiv {\rm IDi}[z]\,,
\end{align}
where ${\rm Erf}[x]$ and ${\rm Di}[x]$ is the Error function and the Dawson integral,\footnote{Explicitly the Dawson integral is given by ${\rm Di}[x]=e^{-x^2}\int_0^x e^{t^2}dt.$
} respectively.
For general values of $\alpha>1$, we find accurate analytical approximations via a combination of a Taylor expansion and a matching to the steepest descent approximation for large arguments. We explain the procedure in App.~\ref{app:asymptoticcalE}. We use this approach, cross-checked with numerical integration, to later evaluate all the integrals in ${\cal I}^{(m)}_j$ and ${\cal I}^{(m)}_j$. All the details of the formulas and approximations can be found in App.~\ref{app:exponentialintegral}.

\begin{figure}
\includegraphics[width=0.49\columnwidth]{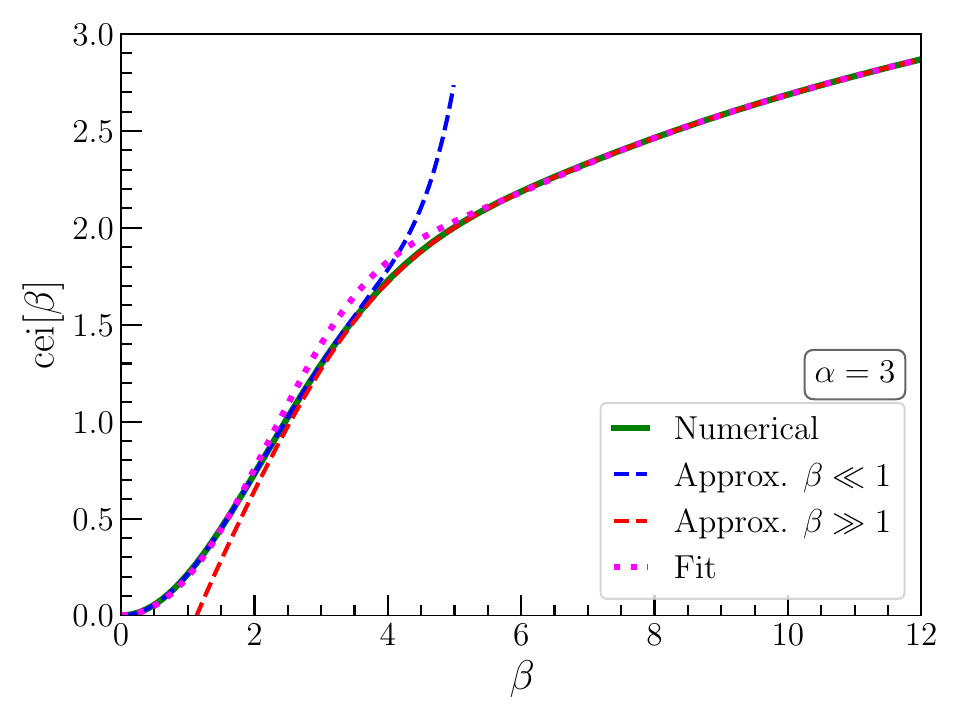}
\includegraphics[width=0.49\columnwidth]{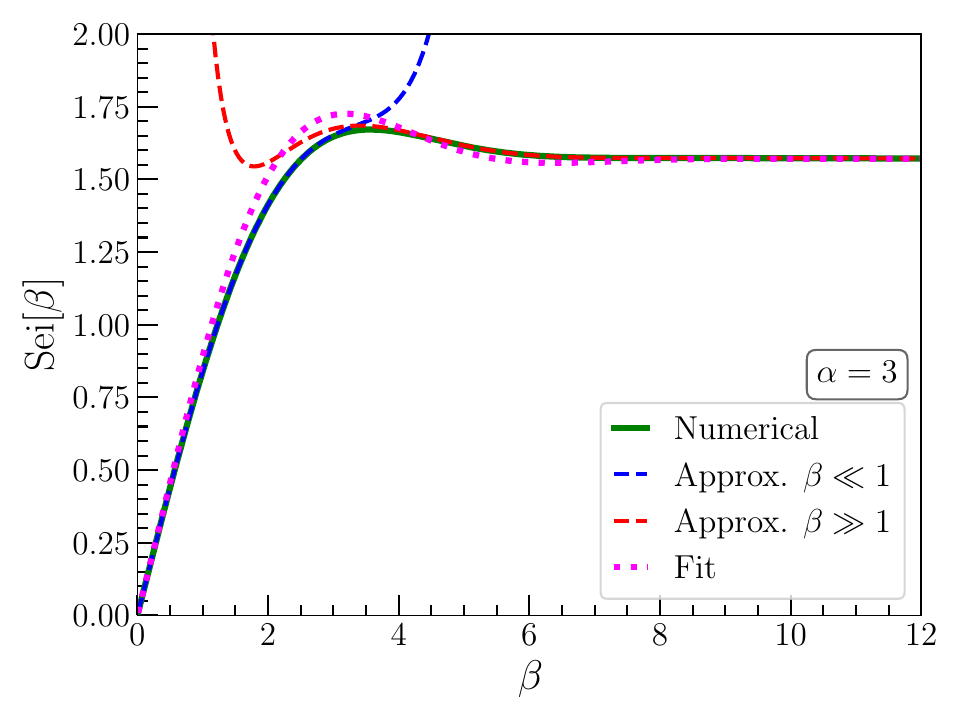}
\caption{The integrals ${\rm cei}[\beta]$ (on the left) and ${\rm Sei}[\beta]$ (on the right) from Eq.~\eqref{eq:cei012} for $\alpha=3$ as a function of $\beta$. Dotted magenta lines show the empirical fits \eqref{eq:fitcei} and \eqref{eq:fitsei}. Dashed lines show the analytical approximations of Eqs.~\eqref{eq:ceiseiapplowb} (in blue, $\beta\ll1$) and \eqref{eq:ceiseilargeb} (in red, $\beta\gg1$) respectively. Note that in Eq.~\eqref{eq:ceiseiapplowb}, we cut the Taylor expansion at $m=10$, which means $5$ terms in the Taylor expansion for ${\rm cei}$ and $5$ for ${\rm Sei}$. See how we can get an accurate approximation by matching the analytical asymptotic limits \eqref{eq:ceiseiapplowb} and \eqref{eq:ceiseilargeb}. We also see that the empirical fit is accurate at the level of $10\%$ at around $\beta\sim 3$ but correctly recovers the asymptotic behaviour. \label{fig:ceiandSei3}}
\end{figure}

\begin{figure}
\includegraphics[width=0.49\columnwidth]{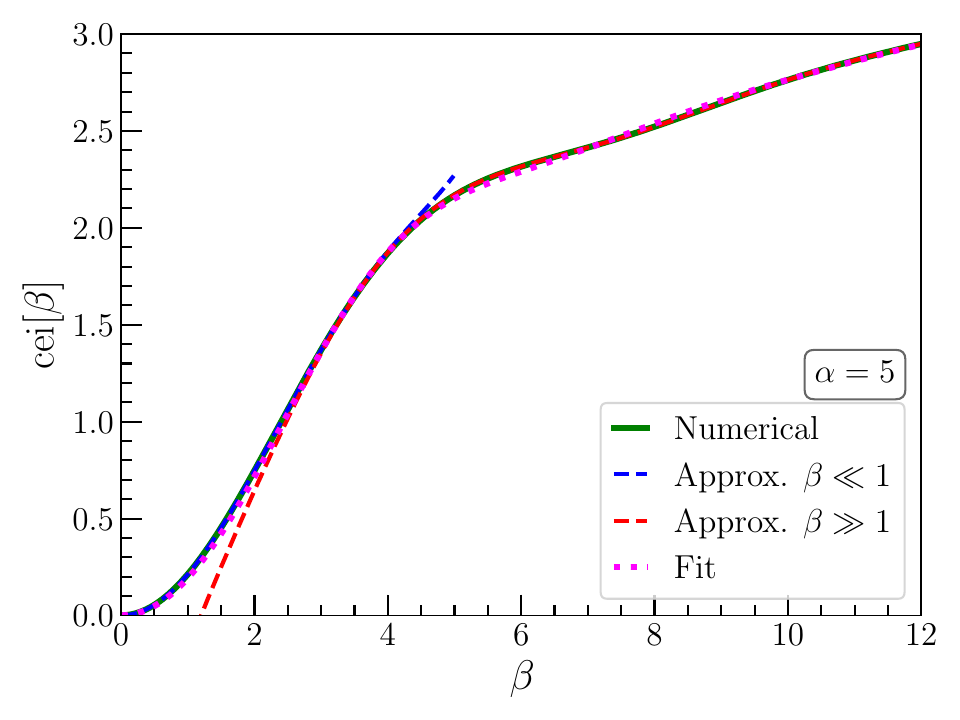}
\includegraphics[width=0.49\columnwidth]{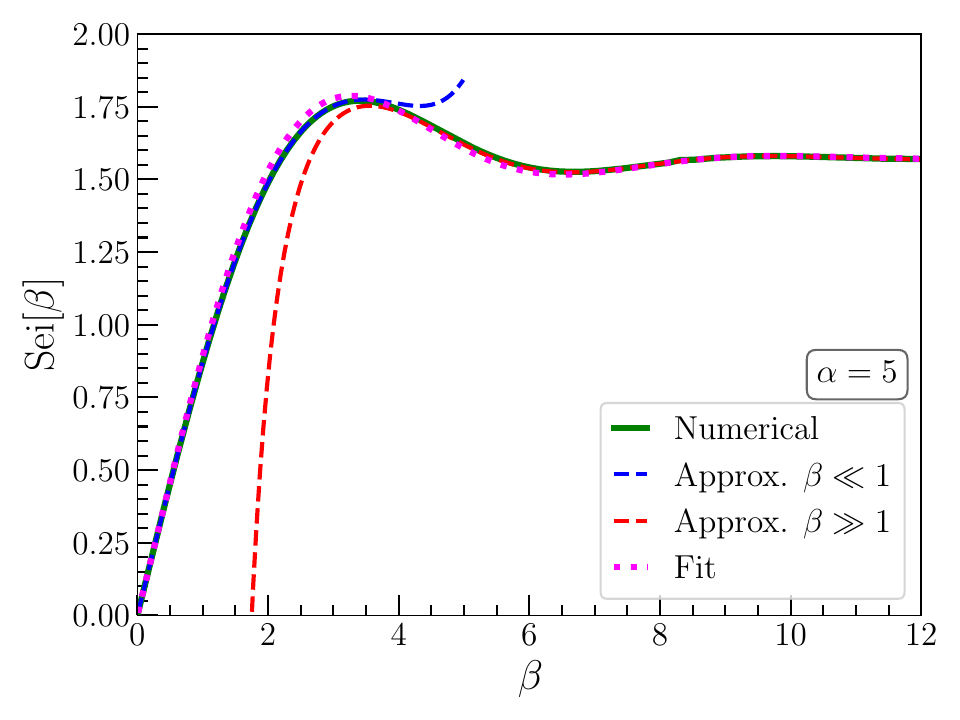}
\caption{As in Fig.~\ref{fig:ceiandSei3}, we show the integrals ${\rm cei}[\beta]$ (on the left) and ${\rm Sei}[\beta]$ (on the right) from Eq.~\eqref{eq:cei012} for $\alpha=5$ as a function of $\beta$. Dotted magenta lines show the empirical fits, Eqs.~\eqref{eq:fitsei} and \eqref{eq:fitcei}, and dashed lines show the analytical approximations of Eqs.~\eqref{eq:ceiseiapplowb} (blue) and \eqref{eq:ceiseilargeb} (red). Again, we can get an accurate approximation by matching the analytical asymptotic limits \eqref{eq:ceiseiapplowb} and \eqref{eq:ceiseilargeb}. Now, we also see that the empirical fit is accurate at the level of less than $1\%$. Empirical fits work best for large values of $\alpha$. \label{fig:ceiandSei5}}
\end{figure}

For $\alpha>2$ some integer and fractional values of $\alpha$ can be written in terms of Hypergeometric functions,\footnote{The case $\alpha=3$ is closely related to the Airy functions. However, we found no way to express it in terms of Airy functions alone as Hypergeometric functions appear as well.} which unfortunately do not appear to be very useful to us. Instead, we provide a fitting formula with a combination of the analytical asymptotic approximations and numerics. Also, we take into account that in the limit of $\alpha\to \infty$ the integrals should be well approximated by placing an upper cut-off in the integrals, resulting in standard sine and cosine integrals. Thus, we propose the following fit for $\alpha>2$ by smoothly patching asymptotic behaviours with decaying exponentials. First, we have that
\begin{align}\label{eq:fitsei}
{\rm Sei}[\beta,\alpha>2]\approx\left(1-e^{-{\beta^2}/{\beta_{s}^2}}\right){\rm sign}[\beta]\frac{\pi }{2}+e^{-{\beta^2}/\beta_{s}^2}{\rm Si}[\sigma_s\beta]\,,
\end{align}
where $\sigma_s=\Gamma[1+\alpha^{-1}]$ and $\beta_{s}\approx 1+1.11\alpha$, the latter a numerical fit for $\alpha\in[3,7]$.
Then, we have
\begin{align}\label{eq:fitcei}
{\rm cei}[\beta,\alpha>2]\approx\left(1-e^{-{|\beta|^3}/{\beta_{c}^3}}\right)& \left( 
   \left(1-\alpha^{-1}\right)\gamma_E+\log|\beta|\right)\nonumber\\&-e^{-{|\beta|^3}/{\beta_{c}^3}}
   ({\rm Ci}|\sigma_{c}\beta|-\log|\sigma_{c}\beta|-\gamma_E )\,,
\end{align}
where $\sigma_{c}=\Gamma[1+2\alpha^{-1}]$ and $\beta_{c}\approx 2+0.66\alpha$, again fitted numerically for $\alpha\in[3,7]$. The fit has a maximum relative error of $10\%$ for $2<\alpha\lesssim3$ and $\beta\sim {\cal O}(1)$, but the fit becomes increasingly better for larger values of $\alpha$. Note that by construction the fit recovers the correct asymptotic limits. We show our fits \eqref{eq:fitcei} and \eqref{eq:fitsei}, as well as the analytic asymptotic expressions of Eqs.~\eqref{eq:ceiseiapplowb} and \eqref{eq:ceiseilargeb}, together with numerical computation in Figs.~\ref{fig:ceiandSei3} and \ref{fig:ceiandSei5}. See how the asymptotic expressions \eqref{eq:ceiseiapplowb} and \eqref{eq:ceiseilargeb} yield a very good approximation. We also see that the fits \eqref{eq:fitcei} and \eqref{eq:fitsei} become increasingly better for larger values of $\alpha$. Note that although we have focused on integer values of $\alpha\geq1$, all formulas are also valid for non-integer values.

\subsection{Analogy with the finite life-time of atomic energy levels\label{sec:analogy}} 

As a small detour, let us construct an interesting analogy with the emission of photons from atomic energy levels. First, note that the two-photon emission spectra for the hydrogen atom have, in some regions, a similar shape to the induced GW spectrum (see, e.g., Fig.~2 of Ref.~\cite{Chluba:2007qk}). Namely, in comparison to the one-photon Lorentzian profiles arising from first-order perturbation theory in quantum mechanics, the two-photon profiles have interference minima and divergent frequencies when directly applying the second-order QM treatment originally developed by Maria G\"oppert-Mayer in her PhD thesis in 1931 \citep{GM1931AnP...401..273G}. The divergences in these photon cross-sections are ``regularized'' by taking into account the finite lifetime of the transition. We expect that a similar interpretation also applies to the induced GW kernel. 

To see that, let us rewrite the average squared kernel \eqref{eq:generalIkernel} in a more compact form by using the complex kernel
\begin{align}
{\cal I}^{(0)}={\cal I}^{(0)}_j+i{\cal I}^{(0)}_y\,.
\end{align}
It is easy to see that the averaged squared kernel \eqref{eq:generalIkernel} is related to the modulus of ${\cal I}^{(0)}$, explicitly by
\begin{align}
2x^2\overline{I^2(u,v,x)}=\left({\cal I}^{(0)}_j\right)^2+\left({\cal I}^{(0)}_y\right)^2=\left|{\cal I}^{(0)}\right|^2\,.
\end{align}
Using the expressions for ${\cal I}^{(0)}_j$ and ${\cal I}^{(0)}_y$, Eqs.~\eqref{eq:calIj0} and \eqref{eq:calIy0}, we find that
\begin{align}\label{eq:I0definition}
{\cal I}^{(0)}=-\frac{y^2}{2c_s^2 u v}\left\{\frac{2}{y}+{\rm Gei}[\beta y_1]+{\rm Gei}[\beta y_2]-{\rm Gei}[\beta y_3]-{\rm Gei}[\beta y_4]\right\}\,,
\end{align}
where $\beta$ is given by Eq.~\eqref{eq:betaj} and we have introduced
\begin{align}
y\equiv\sqrt{1+\frac{y_1y_2y_3y_4}{(2c_s^2 u v)^2}}=
\frac{1-c_s^2 \left(u^2+v^2\right)}{2 c_s^2 u v}\,,
\end{align}
for compactness. We also defined
\begin{align}\label{eq:Geii}
{\rm Gei}[\beta y_j]=\int_{\epsilon}^\infty \frac{d z}{ z} e^{- z^\alpha-i\beta y_j z}\,,
\end{align}
where $\epsilon$ has the role of a regulator. Note that the complex kernel ${\cal I}^{(0)}$ \eqref{eq:I0definition} is finite when $\epsilon\to 0$.

Now, notice that for $\alpha=1$ the integral ${\rm Gei}[\beta y_j]$ is nothing but an exponential integral, namely
\begin{align}\label{eq:Geii1}
{\rm Gei}[\beta y_j]={\rm E}_1[i\epsilon \xi_j]=\int_{\epsilon}^\infty \frac{d z}{ z} e^{-i\xi_j z}\quad{\rm with}\quad \xi_j=y_j-i/\beta\,.
\end{align}
In the limit $\epsilon\to 0$, we have that ${\rm E}_1[z]\to -\ln z-\gamma_E$. Thus, for $\alpha=1$, we have that Eq.~\eqref{eq:I0definition} becomes
\begin{align}\label{eq:I0alpha1}
{\cal I}^{(0)}\approx \frac{y^2}{2c_s^2 u v}\left(\frac{2}{y}+\ln\frac{\xi_3\xi_4}{\xi_1\xi_2}\right)=\frac{y^2}{2c_s^2 u v}\left[\frac{2}{y}+\ln\frac{(\beta y_3-i)(\beta y_4-i)}{(\beta y_1-i)(\beta y_2-i)}\right]\,.
\end{align}
Note that we recover Eqs.~\eqref{eq:seiandcei} when selecting the real and imaginary parts of Eq.~\eqref{eq:I0alpha1}. When $\beta\to\infty$, the imaginary part appears whenever we have $y_3<0$, picking up a factor of $i\pi$ from the logarithm.

If we now guration where $1-c_s(u+v)=\delta\ll1$, then the kernel is at leading order in $\delta$ approximately given by
\begin{align}\label{eq:I0approx}
{\cal I}^{(0)}(c_s(u+v)\sim 1)\approx \frac{1}{2c_sv(1-c_s v)^3}\ln\left[\frac{(\delta-i/\beta)}{2c_sv(1-c_sv)}\right]\,,
\end{align}
where we used that $y\sim 1/(1-c_sv)$, $y_4\sim 2$, $y_3\sim \delta$, $y_1\sim 2c_sv$ and $y_2\sim 2(1-c_sv)\equiv 2/y$. From Eq.~\eqref{eq:I0approx}, we see that the factor $i/\beta$ basically regularizes the kernel. This means that for large enough $\beta$, the position of the peak in the kernel remains mostly at the resonant configuration $1-c_s(u+v)$ but with a finite amplitude.

Interestingly, in this formulation, we see that can take the well-known results with no damping (i.e. $\beta\to\infty$) and include the damping by introducing an imaginary component to $y_j$. Namely we replace $y_j\to y_j-i/\beta$ or, in terms of $u$ and $v$, $1\pm c_s(u\pm v)\to 1\pm c_s(u\pm v)-i/\beta$. In this way, and with the analogy of the finite lifetime of atomic levels in mind, we can interpret $\beta$ as the ``lifetime'' of the resonance. Indeed, for $x=\beta$, the exponential damping is substantial, and the resonance effectively stops. Unfortunately, we found no way of rigorously generalizing this interpretation to arbitrary values of $\alpha$. However, we note that when $\beta y_j\gg1$, the integral \eqref{eq:Geii} is dominated by the $z\ll 1$ region, effectively removing the effect of $z^\alpha$ at leading order for $\alpha>1$. And, note that $\beta$, defined in Eq.~\eqref{eq:betaj}, already contains a non-trivial $\alpha$-dependence.  Thus, in the regime $\beta y_j\gg1$, it is still a good approximation to replace $y_j\to y_j-i/\beta$ for any value of $\alpha>1$. We will confirm this expectation later in Sec.~\ref{sec:spectrum}.

\subsection{Corrections from derivatives of the exponential damping \label{sec:corrections}}

The explicit expressions for ${\cal I}^{(m)}_j$ and ${\cal I}^{(m)}_y$ in Eq.~\eqref{eq:calIjIy}, that is the coefficients and the integrals, depend on the value of $\alpha$ (see App.~\ref{app:detailskernel}). Unfortunately, we found no smart way to simplify and unify the expressions by integration by parts for an arbitrary value of $\alpha$. Nevertheless, given a fixed value of $\alpha$ one can reduce the integrals to
\begin{align}\label{eq:Gei}
{\rm Gei}_n[y_j]=\int_{0}^\infty {dx}\, x^n\,e^{iy_jx -{(u^2+v^2)\kappa_{D}^2}(x/x_*)^\alpha}\,.
\end{align}
where $n>0$. For the case $n=-1$ we use the definition in Eq.~\eqref{eq:Geii}, namely ${\rm Gei}[y_j]\equiv{\rm Gei}_{-1}[y_j]$. It is straightforward to see that
\begin{align}
{\rm Re}[{\rm Gei}_n[y_j]]={\rm Cei}_{-n}^0[y_j] \quad{\rm and}\quad {\rm Im}[{\rm Gei}_n[y_j]]={\rm Sei}_{-n}^0[y_j]\,,
\end{align}
where ${\rm Re}[x]$ and ${\rm Im}[x]$ stand for real and imaginary parts and the generalized sine and cosine integrals ${\rm Sei}_n^0[y_j]$ and ${\rm Cei}_n^0[y_j]$ are defined in Eqs.~\eqref{eq:Ceigeneral} and \eqref{eq:Seigeneral}.  Below, we will roughly show the leading order asymptotics and refer the reader to App.~\ref{app:exponentialintegral} for all the details. In App.~\ref{app:otherpowerlaw}, we provide the explicit analytical expressions for the exact cases of $\alpha=1$ and $\alpha=2$. 

Let us study the general asymptotic behaviour of ${\rm Gei}_n[y_j]$. We first use the same change of variables as in Eq.~\eqref{eq:cei012} and introduce the factor $\beta$ \eqref{eq:betaj}. Using the results of App.~\ref{app:exponentialintegral}, we find 
\begin{align}\label{eq:geiapprox}
{\rm Gei}_n[\beta y_j]\approx\left\{
\begin{aligned}
&\beta^{1+n}\times {\rm constant}+{\cal O}(\beta^{1+n}(\beta y_j))\quad &(\beta y_j\ll1)\\
&y_j^{-1-n}+{\cal O}(\beta^{1+n}(\beta y_j)^{-\alpha})\quad &(\beta y_j\gg1)
\end{aligned}
\right.\,.
\end{align}
From Eq.~\eqref{eq:geiapprox} we conclude that the correction terms proportional to $\gamma_*$ in Eq.~\eqref{eq:calIjIy} are subdominant, or at most comparable, to ${\cal I}^{(0)}$ \eqref{eq:I0definition} in the asymptotic limits. For instance, for $\beta y_j\ll1$, the leading order term in ${\cal I}^{(0)}$  is proportional to $\beta$, see e.g. Eqs.~\eqref{eq:ceiasympt} and \eqref{eq:Seiasympt}, while ${\rm Gei}_n[\beta y_j]\sim \beta^{1+n}$.  On the other hand, for $\beta y_j\gg 1$, the leading order in ${\cal I}^{(0)}$ is $\ln|\beta y_j|$, while ${\rm Gei}_n[\beta y_j]\sim y_j^{-1-n}$. Thus, for $n>0$, the asymptotic behavior of ${\cal I}^{(m)}_j$ and ${\cal I}^{(m)}_y$ is subleading compared to ${\cal I}^{(0)}$, in addition to being suppressed by factors $\gamma_*^{2m}$. Note that, although these contributions could become relevant for $\beta y_j\sim 1$, we numerically find that this is not always the case in the total kernel. And, if they do, they have the same asymptotic behaviour as ${\cal I}^{(0)}$.

\begin{figure}
\includegraphics[width=0.49\columnwidth]{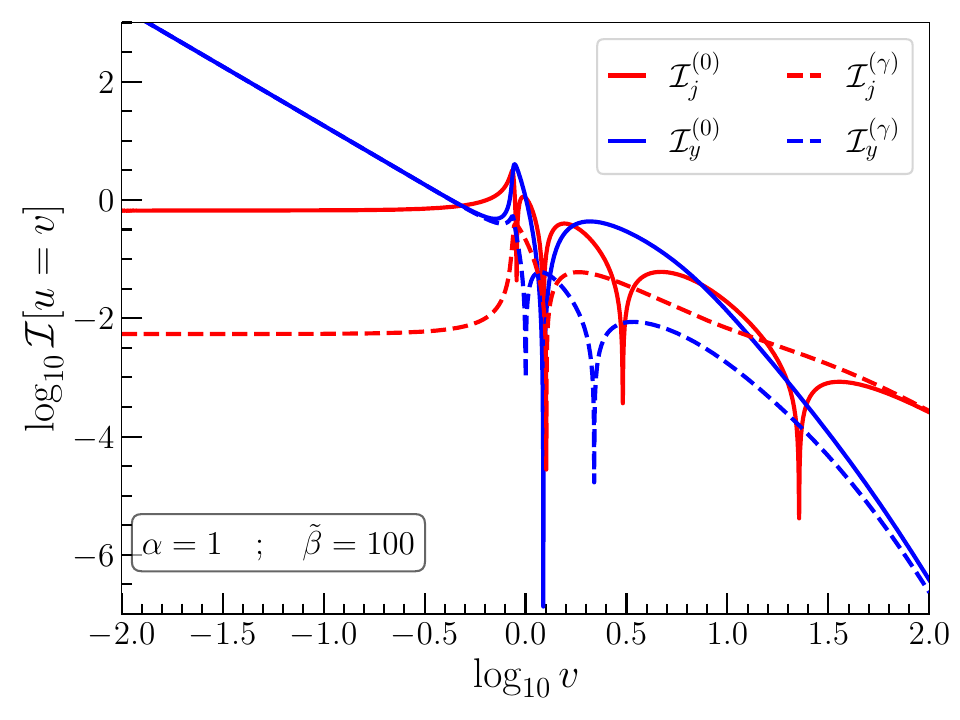}
\includegraphics[width=0.49\columnwidth]{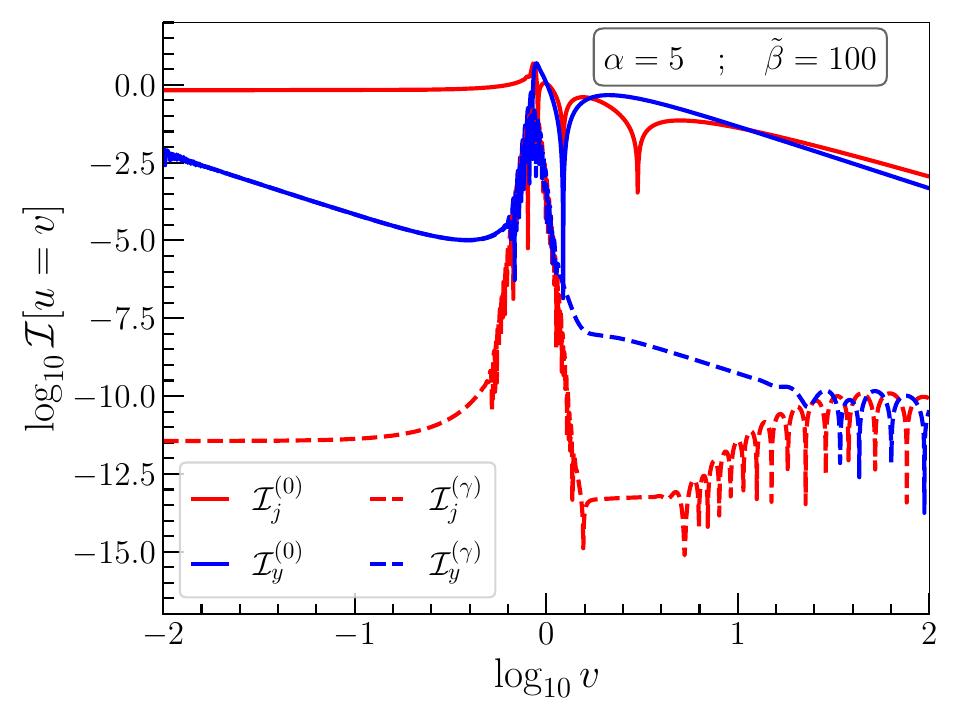}
\caption{Contributions to the kernel $I_j$ (red lines) and $I_y$ (blue lines) as defined in Eq.~\eqref{eq:calIjIy} for $\alpha=1$ (left panel) and $\alpha=5$ (right panel). ${\cal I}^{(0)}_j$ and ${\cal I}^{(0)}_y$ are given by Eqs.~\eqref{eq:calIj0} and \eqref{eq:calIy0} and shown in solid lines. With dashed lines we show the remaining terms proportional to $\gamma_*$ in Eq.~\eqref{eq:calIjIy}, the explicit expression of which follows from the formulas in App.~\ref{app:detailskernel}. See how, in most regimes, the contribution from the correction terms is negligible. In the asymptotic regimes, the contributions can, in some instances, be similar. Note that the contribution from the corrections becomes more suppressed when increasing the value of $\tilde\beta$. We also see that the corrections are more important for $\alpha=1$ than for $\alpha=5$.  \label{fig:kernel}}
\end{figure}

We investigate numerically the contributions from ${\cal I}^{(m)}_j$ and ${\cal I}^{(m)}_y$ and show the results in Fig.~\ref{fig:kernel} for the physically relevant cases of $\alpha=1$ and $\alpha=5$. For the numerical investigation, we pulled out the $u$ and $v$ dependence in $\beta$ in Eq.~\eqref{eq:betaj}, namely, we split $\beta=\tilde\beta\times (u^2+v^2)^{-1/\alpha}$ and defined
\begin{align}
 \tilde\beta\equiv\kappa^{1-2/\alpha}\times \gamma_*^{-2/\alpha}\,.
 \end{align}
Then, in Fig.~\ref{fig:kernel} we fixed $\tilde\beta=100$, which is equivalent to fixing the value of $\kappa$.  We then fixed $u=v$ for simplicity but we have checked that the Kernel behaves similarly at the boundaries of the integration plane, at $u=1+v$ and $u=|1-v|$. Also we find that increasing/decreasing the value of $\tilde\beta$, suppresses/rises the relative contribution of ${\cal I}^{(m)}_j$ and ${\cal I}^{(m)}_y$, but preserves the far asymptotic behavior. For a peaked primordial spectrum, $v\gg1$ contributes to the low-frequency tail of the induced GWs, while $v\ll1$ to the high-frequency tail. From Fig.~\ref{fig:kernel}, we see that the high-frequency tail is the most sensitive to the $\gamma_*$ corrections. But, for very sharp peaks in the primordial spectrum, the high-frequency tail of the induced GW spectrum is very quickly suppressed (see, e.g., Refs.~\cite{Xu:2019bdp,Pi:2020otn,Liu:2020oqe,Atal:2021jyo,Balaji:2022dbi}). Noting that the $\gamma_*$ corrections are subdominant for $v\gg 1$ (or in the case of $\alpha=1$ take some time to approach the asymptotics of ${\cal I}^{(0)}_{j/y}$) and that $\tilde\beta$ increases with decreasing $\kappa$, we conclude that the low-frequency tail of the induced GW spectrum is not sensitive to such corrections. We also confirmed this numerically for the Dirac delta primordial spectrum in Sec.~\ref{sec:spectrum}. We leave the study of the high-frequency tail for relatively broad primordial spectra for future work.

\section{Effects on the induced GW spectrum\label{sec:spectrum}}

In the previous section, we derived the analytical Kernel for induced GWs in the presence of damping. We have seen that the effect of damping is to remove the logarithmic divergence at the resonant configuration. We are, however, more interested in the effects of damping in the induced GW spectrum, see Eq.~\eqref{eq:Phgaussian}, in which the kernel is multiplied by the two primordial spectra integrated over all scalar momentum configurations. If the primordial spectrum has a finite width, we expect a certain smearing effect depending on the width. If the spectrum is broad, the structure of the kernel is basically erased. For example, when no damping is present, a scale-invariant primordial spectrum yields a scale-invariant induced GW spectrum \cite{Kohri:2018awv}. If the spectrum is sharply peaked, some of the structure remains, like the resonant peak and the logarithmic running of the low-frequency tail, but the resonant peak is smoothed out. In order not to mix the effect of finite width and damping, let us consider an infinitely sharp peak, namely
\begin{align}\label{eq:diracdelta}
{\cal P}_\zeta(k)={\cal A}_\zeta\delta(\ln(k/k_*))\,,
\end{align}
where we chose the peak position equal to the pivot scale $k_*$.  We will discuss later the implications of the finite width case.

\subsection{Infinitely peaked primordial spectrum}
In the case of a Dirac delta primordial spectrum, the divergence in the kernel appears in the induced GW spectrum by having an infinitely large peak. One may dismiss the divergence in the spectrum as unphysical because of the zero-width of the primordial spectrum. This argument, however, is unnecessary when considering damping. Let us show that damping leads to a finite amplitude of the induced GW peak. We will focus on the case of a power-law damping for simplicity.  

\begin{figure}
\includegraphics[width=0.49\columnwidth]{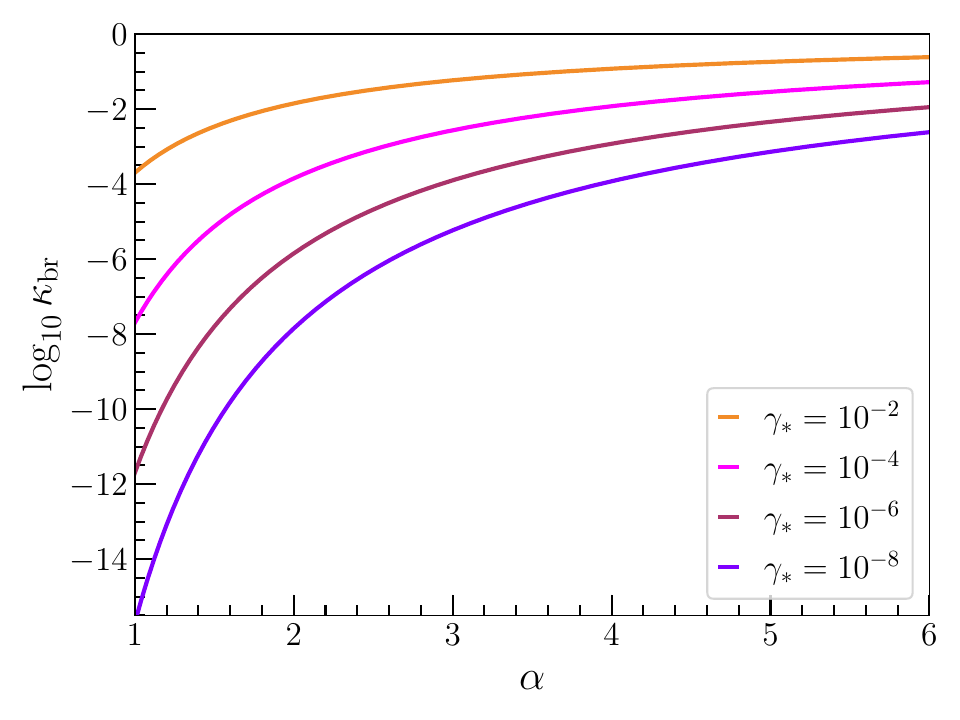}
\includegraphics[width=0.49\columnwidth]{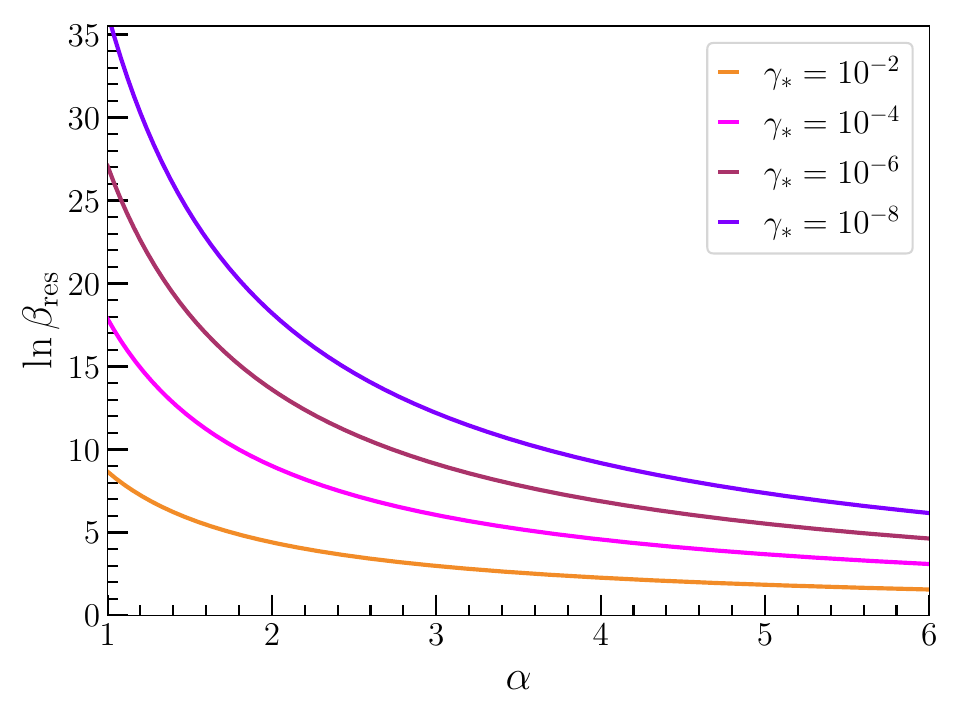}
\caption{On the left panel, we show the breaking scale normalized to the peak scale, $\kappa_{\rm br}=k_{\rm br}/k_*$, given by Eq.~\eqref{eq:kappabr}. The breaking scale determines the breaking of the logarithmic running in the induced GW spectrum, see Eq.~\eqref{eq:IR}. We plot $\kappa_{\rm br}$ as a function of the damping rate index $\alpha$ \eqref{eq:F} for various choices of $\gamma_*$. Specifically, we show $\gamma_*=\{10^{-2},10^{-4},10^{-6},10^{-8}\}$ in orange, magenta, dark red and purple lines, respectively. See how as $\alpha$ increases or $\gamma$ decreases the break scale $\kappa_{\rm br}$ increases. For large values of $\alpha$ the break frequency gets closer to the peak frequency. The purple line shows the lower bound on $\kappa_{\rm br}$ from the standard model of particles \eqref{eq:gammaTex}. On the right panel, we show the damping parameter $\beta$ at the resonant scale, $\beta_{\rm res}=k_{\rm res}/k_{\rm br}$, given by Eq.~\eqref{eq:betares}. The amplitude of the resonant peak in the induced GW spectrum depends on $\ln^2\beta_{\rm res}$, see Eq.~\eqref{eq:peakmax}. The colour code is the same as in the left panel. See how as $\alpha$ increases or $\gamma_*$ decreases, the amplitude of the resonant peak decreases. The purple line shows the upper bound from the standard models of particles.
\label{fig:betakappabr}}
\end{figure}

In the Dirac delta case \eqref{eq:diracdelta}, the momentum integral \eqref{eq:Phgaussian} selects only $u=v=k_*/k=\kappa^{-1}$, and the induced GW spectrum \eqref{eq:spectraldensity} is then given by
\begin{equation}\label{eq:spectraldensitydirac}
\Omega_{\rm GW}(\kappa)=\frac{1}{3}{\cal A}_\zeta^2\kappa^{-2}\left(1-\frac{\kappa^2}{4}\right)^2\left[I_j^2(\kappa)+I_y^2(\kappa)\right]\Theta(2-\kappa)\,,
\end{equation}
where momentum conservation leads to a cut-off at $\kappa=2$. In general, $I_j$ and $I_y$ are given by \eqref{eq:IjIygamma}. However, based on the discussions of Sec.~\ref{sec:corrections} and further numerical checks for the case of Dirac delta, we find that the leading contribution is given by $I_j\approx {\cal I}_j^{(0)}$ and $I_y\approx {\cal I}_y^{(0)}$, Eqs.~\eqref{eq:calIj0} and \eqref{eq:calIy0} respectively. Lastly, we have that the parameter $\beta$ \eqref{eq:betaj}, which describes the magnitude of the damping for a given $\kappa$, simplifies to
\begin{align}\label{eq:bjpeak}
\beta =\frac{\kappa}{(\sqrt{2}\gamma_*)^{2/\alpha}}\equiv \frac{\kappa}{\kappa_{\rm br}}=\frac{k}{k_{\rm br}}\,.
\end{align}
In Eq.~\eqref{eq:bjpeak} we introduced, for later use, the breaking scale
\begin{align}\label{eq:kappabr}
\kappa_{\rm br}= (\sqrt{2}\gamma_*)^{2/\alpha}\,,
\end{align}
which divides the $\beta\ll1$ and $\beta\gg1$ regimes, eventually related to the breaking of the logarithmic running and the finite-size resonant peak. Note that in general $\kappa_{\rm br}\ll1$. We show the dependence of $\kappa_{\rm br}$ on $\alpha$ and $\gamma_*$, as well as $\beta$ evaluated at the resonant peak (see section below) in Fig.~\ref{fig:betakappabr}.
Let us now study two relevant situations: $(i)$ the resonant peak and $(ii)$ the low-frequency tail of the induced GW spectrum.

\subsubsection{The resonant peak}
The resonant peak in the induced GW spectrum lies at $2c_sk_*=k$ ($y_3=0$) which means $\kappa_{\rm res}=2c_s$. At the resonant peak, we also have that
\begin{align}\label{eq:betares}
y_3=0\quad,\quad y_1=y_2=y_4/2=1\quad{\rm and}\quad \beta [\kappa_{\rm res}]=\frac{\kappa_{\rm res}}{\kappa_{\rm br}}\equiv \beta_{\rm res} \,.
\end{align}
Since $\kappa_{\rm br}\ll1$ we have that $\beta_{\rm res}\gg1$ and constant. Thus, after a Taylor expansion for $\delta \kappa\equiv\beta_{\rm res} (1-\kappa_{\rm res}/\kappa)\ll1$, we obtain that
\begin{align}\label{eq:Ij0nearpeak}
{\cal I}^{(0)}_{j,\rm res}&\approx-4+2\left(2{\rm cei}[\beta_{\rm res}]-{\rm cei}[2\beta_{\rm res}]-\Gamma[1+2\alpha^{-1}]\delta \kappa^2/4\right)+{\cal O}(\delta \kappa^3)\nonumber\\&\approx 2\ln\beta_{\rm res}-4-2\Gamma[1+2\alpha^{-1}]\delta \kappa^2+{\cal O}(\delta \kappa^3)\,,
\end{align}
and
\begin{align}\label{eq:Iy0nearpeak}
 {\cal I}^{(0)}_{y,\rm res}&\approx 2\left(2{\rm Sei}[\beta_{\rm res}]-{\rm Sei}[2\beta_{\rm res}]-\Gamma[1+\alpha^{-1}]\delta \kappa\right)+{\cal O}(\delta \kappa^2)\nonumber\\&\approx {\pi}-2\Gamma[1+\alpha^{-1}]\delta \kappa+{\cal O}(\delta \kappa^2)\,.
\end{align}
Note that the leading order coefficients are effectively independent of the value of $\alpha$. 

\begin{figure}
\includegraphics[width=0.49\columnwidth]{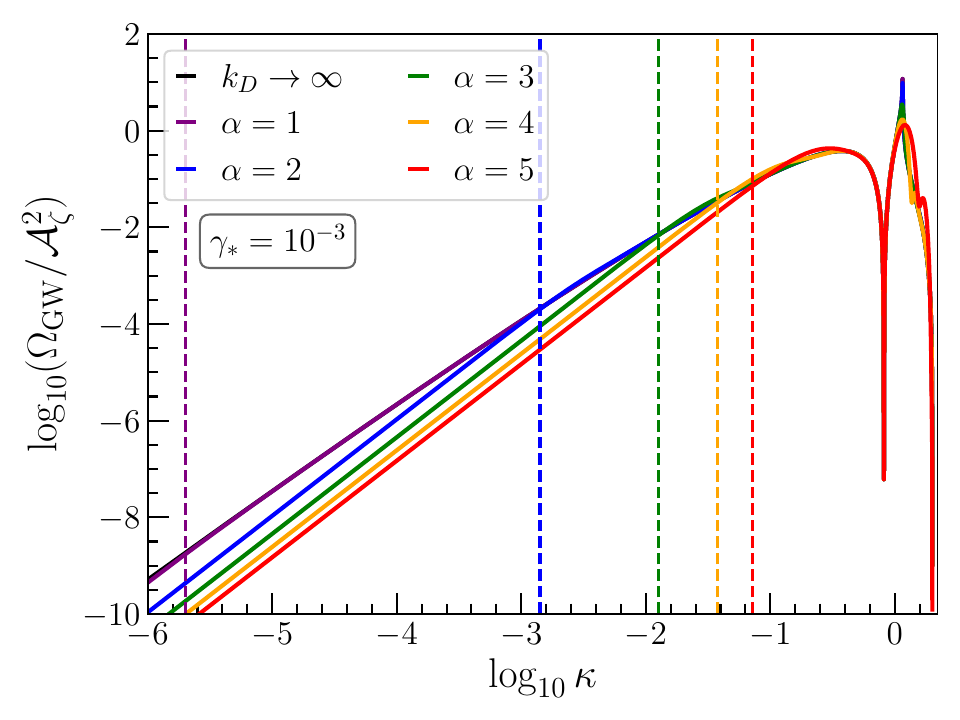}
\includegraphics[width=0.49\columnwidth]{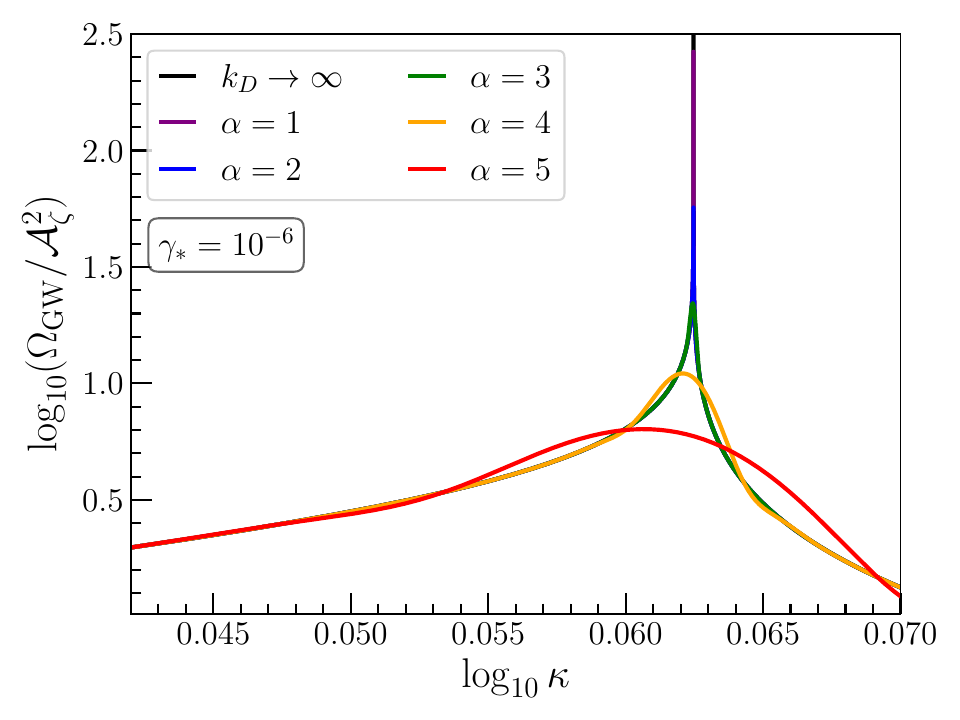}
\caption{Induced GW spectra for a Dirac delta primordial spectrum \eqref{eq:spectraldensitydirac} normalized by ${\cal A}_\zeta^2$ for $\alpha=\{1,2,3,4,5\}$ respectively in purple, blue, green, orange and red lines. In black, we show the case of no damping, for which we explicitly had to remove the pole at the resonance. On the left panel, we show the induced GW spectrum for $\gamma_*=10^{-3}$. With dashed lines, we show the analytical estimate for break frequency \eqref{eq:kappabr} below which the logarithmic running disappears with the same colour code as the solid lines. See how for $\alpha=5$ the logarithmic running is lost already one decade in $k$ below the peak. On the right panel, we zoom in to the resonant peak. We fix $\gamma_*=10^{-6}$, which corresponds to significantly strong damping, to show how the amplitude of the peak depends on $\alpha$. One can clearly see how the divergence is regularized by the damping, showing a profile that is significantly affected for large values of $\alpha$. 
\label{fig:omegas1}}
\end{figure}

From Eqs.~\eqref{eq:Ij0nearpeak} and \eqref{eq:Iy0nearpeak}, we see that the peak of the induced GW spectrum \eqref{eq:spectraldensitydirac} near the resonant frequency in the Dirac delta case is approximately given by
\begin{align}\label{eq:peakmaxef}
\Omega_{\rm GW}^{\delta}(k_{\rm res})&\approx\frac{\left(1-c_s^2\right)^2}{12c_s^2}{\cal A}_\zeta^2\left[({\cal I}^{(0)}_{j,\rm res})^2+({\cal I}^{(0)}_{y,\rm res})^2\right]\nonumber\\&\approx \frac{\left(1-c_s^2\right)^2}{3c_s^2}{\cal A}_\zeta^2\bigg[(\ln\beta_{\rm res}-2)^2+\frac{\pi^2}{4}\nonumber\\&
\qquad-{\pi\Gamma[1+\alpha^{-1}]\delta \kappa-\delta \kappa^2\left(2(\ln\beta_{\rm res}-2)\Gamma[1+2/\alpha^{-1}]-\Gamma[1+\alpha^{-1}]^2\right)}\bigg]\,.
\end{align}
We, therefore, conclude that the peak in the induced GW spectrum is finite and with an amplitude given by
\begin{align}\label{eq:peakmax}
\Omega_{\rm GW}^{\delta,\rm max}(k_{\rm res})&\approx \frac{\left(1-c_s^2\right)^2}{3c_s^2}{\cal A}_\zeta^2\ln^2\beta_{\rm res}\,.
\end{align}
We show this effect clearly in the right panel of Fig.~\ref{fig:omegas1}. From Eq.~\eqref{eq:peakmaxef} we also conclude that in the presence of damping the resonant peak is slighlty shifted towards lower frequencies, because of the linear negative term in $\delta\kappa$, and has a round shape because of the negative $\delta\kappa^2$ term. Note that Eq.~\eqref{eq:peakmax} gives the general maximum that the peak in the induced GW can attain since the resonance is the largest for the Dirac delta case. Also, from the discussions in Sec.~\ref{sec:analogy}, $\beta_{\rm res}$ corresponds to the finite “lifetime” of the resonance.

It is interesting to put some numbers in Eq.~\eqref{eq:peakmax}. In Sec.~\ref{sec:induceddamping}, Eq.~\eqref{eq:gammaTex}, we saw that in the very early universe, there is a lower bound on $\gamma_*$. Namely, assuming the standard model of particles we have that $\gamma_*>10^{-8}$. This means that there is an upper bound on $\beta_{\rm res}$. In particular, we have that
\begin{align}\label{eq:betaresbound}
\ln\beta_{\rm res}<\left\{
\begin{aligned}
&36 &(\alpha=1)\\
&8 &(\alpha=5)
\end{aligned}
\right.\,,
\end{align}
where we chose two representative values of $\alpha=1$ and $\alpha=5$, corresponding to gauge interactions and massive mediators respectively. We also took $c_s=1/\sqrt{3}$. For the bound \eqref{eq:betaresbound} for general $\alpha$ see the purple line in the right panel of Fig.~\ref{fig:betakappabr}. Thus, we see that the amplitude of the resonant peak is in general bounded by
\begin{align}\label{eq:peakmaxsm}
\Omega_{\rm GW}^{\delta,\rm max}(k_{\rm res})&< {\cal A}_\zeta^2\times\left\{
\begin{aligned}
&600 &(\alpha=1)\\
&24 &(\alpha=5)
\end{aligned}
\right.\,.
\end{align}
So, depending on $\alpha$, the resonant peak can only go from ${\cal O}(10)-{\cal O}(100)$ times ${\cal A}_\zeta^2$.

\subsubsection{The logarithmic running of the low-frequency tail}

\begin{figure}
\includegraphics[width=0.49\columnwidth]{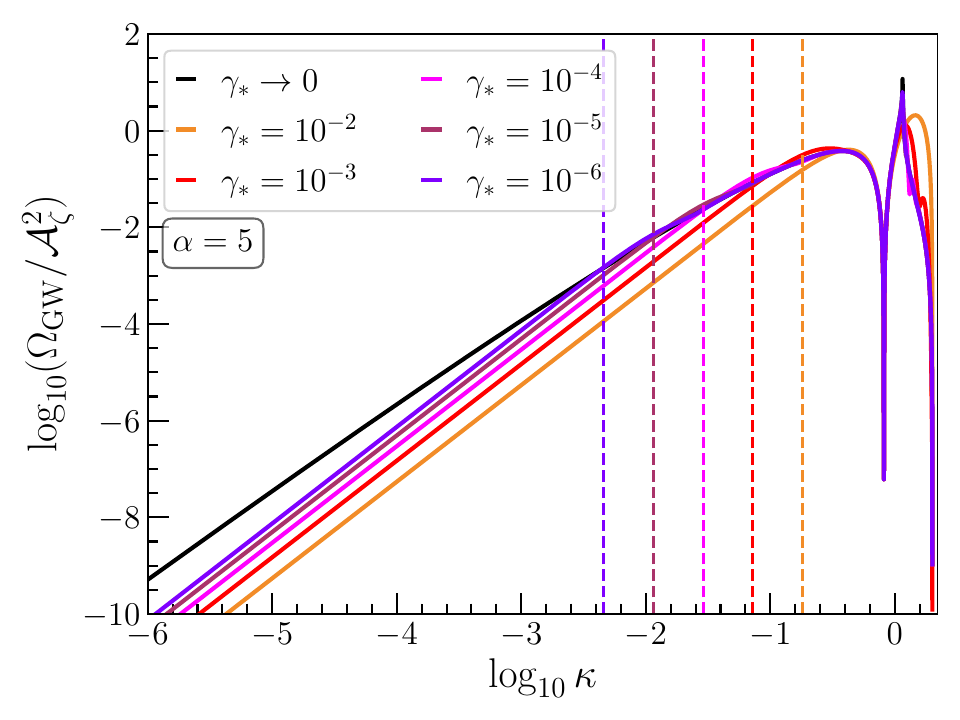}
\includegraphics[width=0.49\columnwidth]{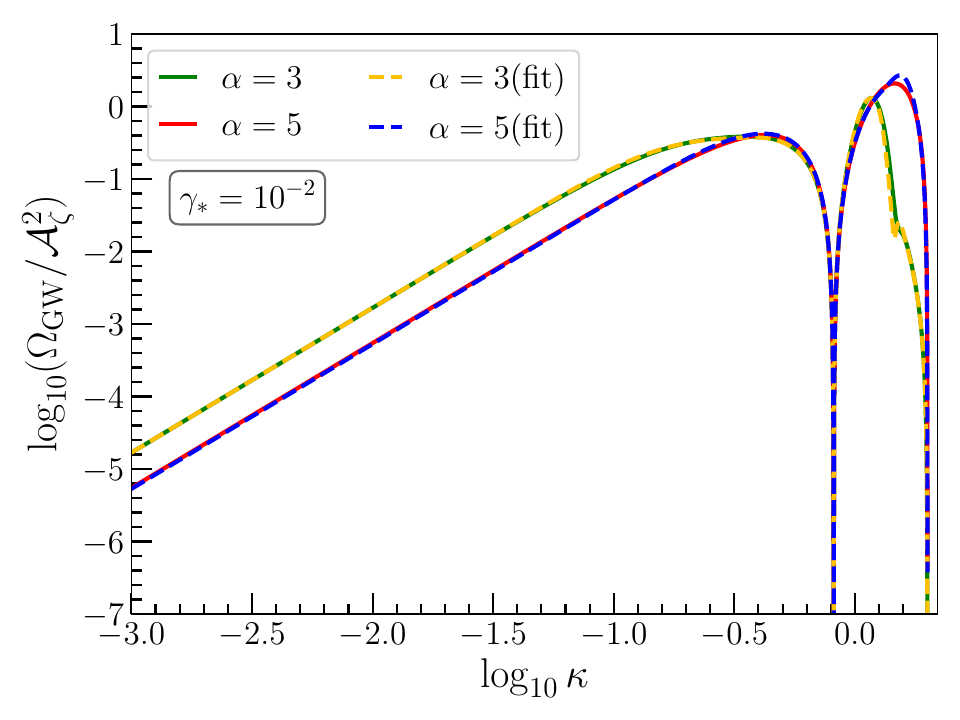}
\caption{Induced GW spectra for a Dirac delta primordial spectrum \eqref{eq:spectraldensitydirac} normalized by ${\cal A}_\zeta^2$. On the left, we fix for $\alpha=5$ and vary $\gamma_*$. We show $\gamma_*=\{10^{-2},10^{-3},10^{-4},10^{-5},10^{-6}\}$ in orange, red, magenta, dark red and purple, respectively. The black line corresponds to no damping or, equivalently, $\gamma_*\to 0$. With dashed coloured lines, we show the value break frequency \eqref{eq:kappabr} for the corresponding value of $\gamma_*$. See how as $\gamma_*$ decreases the resonant peak starts to decrease and eventually becomes a bump. On the right panel, we compare the accurate analytical results (solid green and red lines for $\alpha=3,5$) for the induced GW spectra with the induced GW spectra using the empirical fits (dashed orange and dashed blue for $\alpha=3,5$), given by Eqs~\eqref{eq:fitcei} and \eqref{eq:fitsei}. We fix $\gamma_*=10^{-2}$ for illustration purposes. Difference become less appreciable, the smaller the value of $\gamma_*$. 
\label{fig:omegas2}}
\end{figure}

Let us look now into the low-frequency tail, or equivalently the far peak region, which corresponds to $\kappa\ll 1$. In that regime, we have that
\begin{align}
y_4=-y_3\approx 2c_s/\kappa\quad,\quad y_1=y_2=1\,.
\end{align}
Since ${\kappa}_{\rm br}\ll1$ we see that $\beta y_4=-\beta y_3=\beta_{\rm res}\gg1$ and constant. However, contrary to the resonant case, here we have two regimes, $\beta \gg1$ and $\beta \ll1$, which distinguish a far-low frequency tail from a mid-low frequency tail, respectively. 
Taking this observation into account, the limits of the Kernels in the low-frequency read
\begin{align}
{\cal I}^{(0)}_{j,\rm IR}\approx\frac{\kappa^2}{c_s^2}\left(1+{\rm cei}[\beta ]-{\rm cei}[\beta_{\rm res}]\right)\approx\frac{\kappa^2}{c_s^2}\left\{
\begin{aligned}
&\ln(\kappa/\kappa_{\rm br}) &1\gg\kappa\gg \kappa_{\rm br}\\
&-\ln\beta_{\rm res}&\kappa\ll \kappa_{\rm br}
\end{aligned}
\right.
\end{align}
and
\begin{align}
 {\cal I}^{(0)}_{y,\rm IR}\approx \frac{\kappa^2}{c_s^2}{\rm Sei}[\beta ]\approx  \frac{\kappa^2}{c_s^2}\left\{
\begin{aligned}
& \pi/2 &1\gg\kappa\gg\kappa_{\rm br}\\
& \Gamma[1+\alpha^{-1}]\beta &\kappa\ll \kappa_{\rm br}
\end{aligned}
\right.\,,
\end{align}
where we used the notation “IR” to denote the infra-red, low-frequency tail.
Thus, the mid and far low-frequency tail of the induced GW spectrum is given by
\begin{align}\label{eq:IR}
\Omega_{\rm GW,IR}^{\delta}(k\ll k_*)\approx\frac{{\cal A}_\zeta^2}{3c_s^4}\kappa^2\left\{
\begin{aligned}
&\ln^2(\kappa/\kappa_{\rm br}) &1\gg\kappa\gg \kappa_{\rm br}\\
&\ln^2\beta_{\rm res}&\kappa\ll \kappa_{\rm br}
\end{aligned}
\right.
\end{align}
The effects of the breaking scale are clearly shown in the left panels of Figs.~\ref{fig:omegas1} and \ref{fig:omegas2}. See how there is no logarithmic running of the induced GW spectrum in the far low-frequency tail, in contrast to the case when no damping is present \cite{Cai:2018dig,Yuan:2019wwo}.\footnote{Let us remind the reader that the logarithmic running is a special feature of the radiation-dominated universe.  In general, it is not present for arbitrary equation of state of the universe, see Refs.~\cite{Domenech:2019quo,Domenech:2020kqm}.} We also see how the amplitude of the logarithmic tail is related to the amplitude of the resonant peak through $\beta_{\rm res}$. This is because both effects have the same origin, namely, a secular subhorizon growth. Thus, $\beta_{\rm res}$ can also be interpreted as the finite “lifetime” of the low-frequency logarithmic running. And, eventually, as the  “lifetime” of the induced GW source.

It is interesting to study the values of $\kappa_{\rm br}$, related to $1/\beta_{\rm res}$, which tells us how far from the peak we lose the logarithmic running, for the two relevant cases in the standard model. Again, using the lower bound from Eq.~\eqref{eq:gammaTex}, namely $\gamma_*>10^{-8}$, we see that
\begin{align}\label{eq:brSM}
\kappa_{\rm br}>\left\{
\begin{aligned}
&2\times 10^{-16} &(\alpha=1)\\
&7\times 10^{-4} &(\alpha=5)
\end{aligned}
\right.\,.
\end{align}
For the bound \eqref{eq:brSM} for general values of $\alpha$ see the purple line in the left panel of Fig.~\ref{fig:betakappabr}.
Interestingly, we see that for $\alpha=5$, the effect of damping is substantial for the low-frequency tail. In that case, the logarithmic running disappears for $k<10^{-3}k_*$. This is not too far from the peak, about 3 orders of magnitude in frequency. So, it would not be hopeless to detect the breaking frequency for a loud enough induced GW signal. For $\alpha=1$, the effect on the low-frequency tail is negligible for observations unless $\gamma_*$ is sizeable. Nevertheless, recall that the values given in Eq.~\eqref{eq:brSM} are a lower bound, assuming the standard model of particles. If there are new very weakly interactive particles in the early universe, $\kappa_{\rm br}$ can be substantially larger. Thus, finding a break frequency in the induced GW spectrum where the logarithmic running disappears is a direct probe of new weakly interacting particles in the very early universe.

\begin{figure}
\includegraphics[width=0.6\columnwidth]{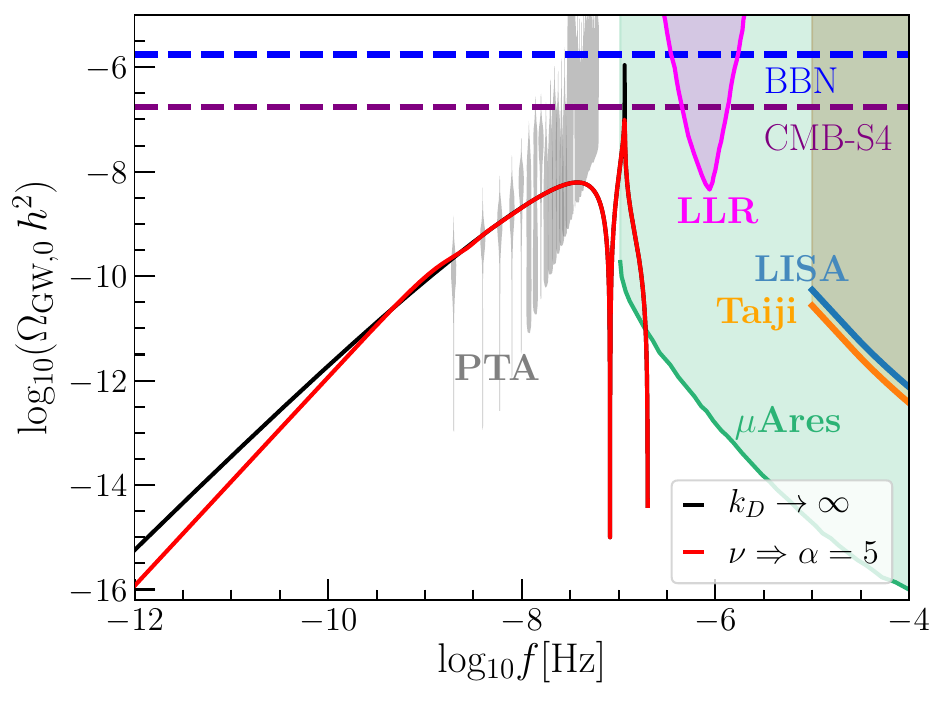}
\caption{Induced GW spectrum from a Dirac delta peak with and without damping (in red and black respectively) as a possible interpretation of the PTA data.  The grey violins indicate the NANOGrav results \cite{NANOGrav:2023hvm}. We also show the power-law integrated sensitivity curves~\cite{Thrane:2013oya} for LISA, Taiji \cite{Barke:2014lsa,Ruan:2018tsw}, $\mu$-Ares \cite{Sesana:2019vho} as well as Lunar Laser Ranging \cite{Blas:2021mqw}, respectively in blue, orange, green and magenta.  To estimate the damping we used the values from the standard model of particles, given by Eq.~\eqref{eq:k_D2}. 
For illustration purposes, we fix $f_*=10^{-7}\,{\rm Hz}$ and $\log_{10}{\cal A}_\zeta=-1.5$. From Eq.~\eqref{eq:gamma}, see also the right panel of Fig.~\ref{fig:kdgamma}, it follows that $\gamma_*\approx 10^{-6}$ and $\alpha=5$. It should be noted that for the black line, the divergence at the resonant frequency is cut due to numerical resolution. In practice, though, one should take into account that observations have a finite bandwidth \cite{Thrane:2013oya,Byrnes:2018txb,Iovino:2024tyg}. When damping effects are included this issue naturally disappears even without intervention.
See how, in the case of a Dirac delta primordial spectrum, the effects of damping are appreciable even in the standard model of particles. Unfortunately, even if we change the position of the peak, the break frequency \eqref{eq:kappabr} is slightly below the nHz frequencies. Nevertheless, this example showcases the importance of considering the damping of fluctuations in the early universe in the induced GW spectrum. 
\label{fig:omeganHz}}
\end{figure}

As an interesting application, we note from the evolution of $\gamma(T)$ in the standard model of particles, see Fig.~\ref{fig:kdgamma}, that the epoch in the early universe with the higher damping rate (with $\alpha=5$) and a growing $\gamma$ happens for $T<100{\rm GeV}$. Since, in the Dirac delta case, $\gamma_*$ is given by $\gamma(T)$ evaluated at the time that the peak $k_*$ enters the Hubble radius, we see that damping could be most important for induced GW peak frequencies ranging from $10^{-5}{\rm Hz}$ to  $10^{-10}{\rm Hz}$. Curiously, this covers the 2-$\sigma$ contours of the Bayesian fit from NANOGrav \cite{NANOGrav:2023hvm} (see left panel of Fig.~7 of \cite{NANOGrav:2023hvm}), assuming the induced GW interpretation of the PTA data. To qualitatively illustrate the importance of damping in the nHz frequency range, without attempting any Bayesian analysis at the moment, we take one value from the 2-$\sigma$ contours of NANOGrav \cite{NANOGrav:2023hvm} ($f_*=10^{-7}\,{\rm Hz}$ and $\log_{10}{\cal A}_\zeta=-1.5$) and plot the resulting spectrum with and without damping from the standard model in Fig.~\ref{fig:omeganHz}. See how, although current PTA data is definitely not sensitive enough, the difference between spectra is visible to the eye. It is thus plausible that, in the future, the effects of damping are important, especially if we have access to the $\mu$Hz regime of the GW background, e.g. with $\mu$-Ares \cite{Sesana:2019vho}, to probe the peak of the induced GW background. Note that other interesting, short-term proposals to test the $\mu$Hz GW background are relative astrometry with photometric surveys \cite{Wang:2022sxn,Pardo:2023cag} (such as Gaia and Nancy Grace Roman Telescope) and the Uranus Orbiter and Probe mission \cite{Zwick:2024hag}.

\subsection{Finite width primordial spectrum}

We end this section with a brief discussion on the case of a finite-width primordial spectrum, e.g. a log-normal or a broken power-law primordial spectrum. Although a detailed analysis is out of the scope of the paper, we can draw general conclusions using previous results in the literature for the case with no damping (see, e.g., Refs.~\cite{Pi:2020otn,Fumagalli:2020nvq}). Let us focus on the sharp log-normal case with logarithmic width $\Delta$, studied extensively in Ref.~\cite{Pi:2020otn}, for simplicity.

First, the effect of a finite width is to smooth out the resonant peak. Analytical approximations in the case of no damping yield that the amplitude of the resonant peak is given by \cite{Pi:2020otn}
\begin{align}\label{eq:omeganodamping}
\Omega_{\rm GW}^{\Delta,\rm res}({\rm no\,\,damping} \sim  \Delta\ll 1/\beta_{\rm es})\approx \frac{1}{2\Delta}\int^{\kappa e^{\Delta}}_{\kappa e^{-\Delta}}d\ln\kappa \,\Omega_{\rm GW}^{\delta,\rm res}(\kappa)\,.
\end{align}
But, note that the width of the resonant peak in the damping case is proportional to $1/\beta_{\rm res}$. We thus see that there will be a competition between the effect of the finite width $\Delta$ and the smoothed resonance $\beta_{\rm res}$ due to damping. In the end, the largest width will dominate. This means that, for $\Delta\times \beta_{\rm res}\gg 1$ the finite width effect is more important, while for $\Delta\times \beta_{\rm res}\ll1$ the damping dominates. In the cases we considered here we have that $\beta_{\rm res}$ is fairly large, so the damping is only important when $\Delta$ is extremely small or the damping is very strong (namely for large values of $\alpha$ and/or sizeable values of $\gamma_*$).

Second, and most interesting, the low-frequency tail will have three different regimes as the finite width introduces a new scale given by $\kappa_{\Delta}\approx 2\Delta$ \cite{Pi:2020otn}. For $\kappa\gg \kappa_\Delta$ one has effectively the Dirac delta result, for $\kappa\ll \kappa_\Delta$ the low-frequency tail decays with an additional power of $\kappa$ faster with decreasing $\kappa$, consistent with causality arguments \cite{Cai:2019cdl}. In the presence of damping, we have the additional scale $\kappa_{\rm br}\,(\sim 1/\beta_{\rm res})$. We now have two possibilities. For $\kappa_{\rm br}>\kappa_\Delta$, the induced GW spectrum transitions first from $k^2\ln^2k$ for $\kappa\ll\kappa_\Delta$ to $k^2$ for $\kappa_\Delta\ll\kappa\ll \kappa_{\rm br}$ and then to $k^3$ for $\kappa\ll \kappa_{\Delta}$. In the opposite case, namely for $\kappa_{\rm br}<\kappa_\Delta$, then it goes from $k^2\ln^2k$ for $\kappa\ll\kappa_{\rm br}$ to $k^3\ln^2k$ for $\kappa_{\rm br}\ll\kappa\ll \kappa_{\Delta}$ and then to $k^3$ for $\kappa\ll \kappa_{\rm br}$. Note that, in the cases we have considered, we generally have low values of $\kappa_{\rm br}$. So, in most situations, we will be in the $\kappa_{\rm br}<\kappa_\Delta$ regime (unless we have large values of $\alpha$ and/or sizeable values of $\gamma_*$). Our results show that induced GWs, when including damping effects, lose the logarithmic running \cite{Cai:2018dig,Yuan:2019wwo} and recover the universal infrared scaling of GW background spectra \cite{Cai:2019cdl}, namely that $\Omega_{\rm GW}\propto f^3$ at very low frequencies.

\section{Discussions and conclusions\label{sec:conclusions}}

The plasma of relativistic particles in the early universe acts like an imperfect fluid with viscosity. Any small-scale density fluctuation below the damping scale becomes exponentially damped due to such dissipative effects. For the GWs induced by primordial density fluctuations, damping of density fluctuations translates into an effective finite ``lifetime'' of the GW generation. For peaked primordial spectra, the lifetime of the induced GW source is roughly $x=k\tau\sim\beta_{\rm res}$ with $\beta_{\rm res}$ given by \eqref{eq:betares}. Or, in simpler terms, since $\beta_{\rm res}\sim 1/\kappa_{\rm br}$, the source to induced GW is effectively shut down at $\tau\sim 1/k_{\rm br}$, where $k_{\rm br}$ is the breaking scale \eqref{eq:kappabr}. As we have shown, the immediate consequence of a finite lifetime of the source is a finite size resonant peak \eqref{eq:peakmax}, and a breaking of the logarithmic running in the low-frequency tail \eqref{eq:IR} (recovering the universal infrared scaling of GW background spectra \cite{Cai:2019cdl}). We find this effect to be especially important for large damping rates, as occurs, e.g., in the standard model of particles below the EW phase transition. The latter concerns induced GWs that enter the nHz window of PTAs. See, e.g., Figs.~\ref{fig:omegas1} and \ref{fig:omeganHz}. 

The absence of the logarithmic running (see, e.g., Fig.~\ref{fig:omegas2}) in the large-scale regime could have important consequences when it comes to the observation of polarization $B$-modes in the CMB with upcoming experiments like the Simons Observatory, {\it Litebird} and CMB-S4. As recently highlighted, existing limits on scalar perturbations still allow a non-negligible contribution from GWs, potentially swamping direct primordial $B$-modes from inflation \cite{Cyr:2023pgw}. With the modified GW spectral template, the contributions from small-scale perturbations relevant to the PTA and interferometer bands should not leak into the CMB bands as strongly. This renders CMB spectral distortion constraints on scalar perturbations at wavenumber $10\,{\rm Mpc^{-1}}\lesssim k \lesssim 10^4\,{\rm Mpc^{-1}}$, see e.g. Ref.~\cite{Chluba2012Inflaton}, more important. The expected improvements over {\it COBE/FIRAS} with CMB spectrometer concepts like {\it FOSSIL} or {\it PIXIE} (for discussion of spectral distortion science see Refs.~\cite{Chluba:2019kpb,Chluba2021ExA}) should therefore allow excluding any scalar-induced GW contributions to the CMB anisotropies. However, we leave a detailed computation to another paper.

The finite ``lifetime'' of the induced GW source may have wider implications for the induced GWs than those studied in this paper. For instance, it may alleviate the gauge issue of the induced GW spectrum \cite{Hwang:2017oxa,DeLuca:2019ufz,Tomikawa:2019tvi,Gong:2019mui,Inomata:2019yww,Yuan:2019fwv}. In Ref.~\cite{Domenech:2020xin} it was shown that one important condition for an approximate gauge independence of the induced GW spectrum is that the source term becomes effectively inactive. Damping achieves precisely this but exponentially faster and, therefore, it might broaden the class of gauges with approximate gauge independence. 

Damping might also be important for the poltergeist mechanism \cite{Inomata:2020lmk} (see also Refs.~\cite{Inomata:2019ivs,Pearce:2023kxp}), where a sudden transition from an early matter domination to a radiation dominated universe enhances the production of induced GWs. The enhancement is due to the fact that density fluctuations grow during the matter-dominated stage and are suddenly transferred to the radiation fluid, which afterwards develops huge acoustic waves. However, the largest induced GW production mainly comes from the resonant momentum configuration \cite{Inomata:2020lmk} and it may take some time from the onset of radiation domination to develop a large induced GW strain. If the finite lifetime of the resonance is shorter than the time for GW production, then damping will be a hindrance to the poltergeist mechanism. We note, however, that in the standard model of particles, the effects are probably small for early matter domination ending above the EW phase transition (see Figs.~\ref{fig:kdgamma} and \ref{fig:omegas1}). However, they could be important for temperatures below the EW scale or in the presence of very weakly interacting beyond standard model particles. We leave a detailed study of the effects of damping on the gauge issue and the poltergeist mechanism for future work.

Before ending, let us note that in our study we have not included the effects of neutrino free streaming \cite{Baumann:2007zm} (or other light particles \cite{Hook:2020phx}) nor the effects of the QCD phase transition \cite{Abe:2020sqb,Franciolini:2023wjm} (in, e.g., Fig.~\ref{fig:omeganHz}). We also limited ourselves to Gaussian primordial fluctuations and sharply peaked primordial spectra. It would be interesting to study the impact of damping on broadly peaked primordial spectrum (such as a scale-invariant spectrum) as well as on the non-Gaussian corrections to the induced GWs \cite{Unal:2018yaa,Cai:2018dig,Atal:2021jyo,Adshead:2021hnm,Abe:2022xur,Garcia-Saenz:2023zue}.

For the reader's convenience, we end by summarizing here our main formulas for the induced GW Kernel.  For the general case, they can be found in Eqs.~\eqref{eq:Ijapprox} and \eqref{eq:Iyapprox} and in App.~\ref{app:detailskernel}. In Sec.~\ref{ref:exactandapproximate}, we provide exact and approximate formulas for the Kernel in the simplified case of power-law damping, which should be sufficient for most applications. It is exciting that we may be able to test for new weakly interacting particles using the low-frequency tail of the induced GW spectrum. We leave a detailed numerical computation in specific models for future work.

\begin{acknowledgments}
GD would like to thank K.~Kohri, M.~Sasaki, G.~Tasinato and A.~Vikman for useful discussions at various stages of the project.  GD is supported by the DFG under the Emmy-Noether program, project number 496592360 and the JSPS KAKENHI grant No. JP24K00624.
\end{acknowledgments}

\appendix

\section{Details on the generalized exponential integrals\label{app:exponentialintegral}}

In this appendix, we derive several properties and asymptotic behaviours of the generalized exponential integrals Eqs.~\eqref{eq:Geii} and \eqref{eq:Gei}. In order to do so, we must first introduce different ways to rewrite the generalized exponential integral Eq.~\eqref{eq:Gei}. One approach will be more useful for deriving recurrence relations, while another for deriving asymptotic approximations.

First, we redefine the integration variable to $x=\beta z$ and introduced
\begin{align}\label{eq:calE}
{\cal E}_n[\beta y_j;\alpha]=\int_0^\infty dz \,z^n e^{-z^\alpha+i\beta y_j z}\,.
\end{align}
Second, we change the integration variables to
\begin{align}\label{eq:xtoz}
x=\left({\beta^\alpha}|y_j|\right)^{\frac{n+1}{\alpha-1}}\tilde z\,,
\end{align}
and introduce
\begin{align}\label{eq:baseE}
E_n[\lambda_j;\alpha]=\int_{0}^\infty {d\tilde z} \,\tilde z^n\,e^{-\lambda_j\left({\tilde z^\alpha}-i\tilde z\right)}\quad{\rm with}\quad \lambda_j=\left(\beta |y_j|\right)^{\frac{\alpha}{\alpha-1}}\,.
\end{align}
Note that for $y_j<0$ we must consider instead its complex conjugate, namely $E^*_n[\lambda_j;\alpha]$.
The relation between these different integrals now reads
\begin{align}
{\rm Gei}_n[y_j]=\beta^{n+1}{\cal E}_n[\beta y_j;\alpha]=\left({\beta^\alpha}|y_j|\right)^{\frac{n+1}{\alpha-1}}E_n[\lambda_j;\alpha]\,,
\end{align}
For completeness we write down explicitly the direct relation between ${\cal E}$ \eqref{eq:calE} and $E$ \eqref{eq:baseE}, which is given by
\begin{align}
{\cal E}_n[\beta y_j;\alpha]=\lambda_j^{\frac{n+1}{\alpha}}E_n[\lambda_j;\alpha]\,.
\end{align}
Note that from now on we will drop the tilde in $z$ in Eq.~\eqref{eq:baseE}. 

\subsection{Properties of the integral \texorpdfstring{${\cal E}_n$}{}}
From the definition \eqref{eq:calE} we note a useful relation when $n\in \mathbb{Z}^+$, namely that
\begin{align}\label{eq:calEnrecurrence}
{\cal E}_{n}[\beta y_j;\alpha]=\frac{1}{(iy_j)^{n}}\frac{\partial^{n}}{\partial\beta^{n}}{\cal E}_{0}[\beta y_j;\alpha]\,.
\end{align}
We also note that
\begin{align}\label{eq:calEntoEm1}
\int_0^{i\beta y_j} d(it)\,{\cal E}_{0}[t;\alpha]={\cal E}_{-1}[\beta y_j;\alpha]-{\cal E}_{-1}[0;\alpha]=i{\rm Sei}[\beta y_j;\alpha]-\,{\rm cei}[\beta y_j;\alpha]\,.
\end{align}
Thus, we only need to study ${\cal E}_{0}[\beta y_j;\alpha]$ to find all the other relevant integrals. When $n\in \mathbb{Z}^+$ there are still some useful relevant relations if we note that in the integrals ${\cal I}_j^{(m)}$ and ${\cal I}_y^{(m)}$ \eqref{eq:calIjIy}, we have that $n=m\alpha-\ell$, where $\ell\in \mathbb{Z}^+$. The useful case to consider is $n=m(\alpha-1)$. We then find that
\begin{align}
{\cal E}_{m(\alpha-1)}=\frac{1}{\alpha}\left[i\beta y_j {\cal E}_{(m-1)(\alpha-1)}+(m-1)(\alpha-1)\frac{\partial{\cal E}_{(m-1)(\alpha-1)}}{\partial(i\beta y_j)}\right]\,.
\end{align}
The recurrence relation stops at
\begin{align}
{\cal E}_{\alpha-1}[\beta y_j;\alpha]=\frac{1}{\alpha}\left(-1+i\beta y_j{\cal E}_{0}\right)\,.
\end{align}
For example, we have that
\begin{align}
{\cal E}_{2\alpha-2}[\beta y_j;\alpha]=\frac{1}{\alpha}\left[i\beta y_j{\cal E}_{\alpha-1}+(\alpha-1){\cal E}_{\alpha-2}\right]=\frac{1}{\alpha^2}\left[(\alpha-1){\cal E}_{0}-i\beta y_j\left(1-i\beta y_j{\cal E}_{0}\right)\right]\,.
\end{align}
These expressions again give all the relevant integrals in terms of ${\cal E}_{0}$.

\subsection{Asymptotic behaviors of \texorpdfstring{${\cal E}_n$}{}\label{app:asymptoticcalE}}
We derive the asymptotic behaviours using $E_n$ \eqref{eq:baseE} in the Sec.~\ref{app:asymptoticEn} but for convenience, we report the corresponding asymptotic behaviours of ${\cal E}_{n}$ here. We focus on the cases with $\alpha>1$ because these are the physically relevant cases but our analysis can be extended to $0<\alpha<1$. First, we find that
\begin{align}\label{eq:calE0smallb}
{\cal E}_{0}[\beta y_j\ll1;\alpha]=  \frac{1}{\alpha}\sum_{m=0}^\infty \frac{\Gamma \left[\frac{1+m}{\alpha }\right]}{m!}(i\beta y_j)^m\,,
\end{align}
and
\begin{align}\label{eq:calE0largeb}
{\cal E}_{0}[\beta y_j\gg1;\alpha]\approx -\frac{1}{i\beta y_j}&\Bigg\{1-\frac{\alpha \Gamma[\alpha]}{(-i\beta y_j)^\alpha}\nonumber\\&
-\sqrt{\frac{2\pi\alpha}{(\alpha-1)}}\left(\frac{i\beta y_j}{\alpha}\right)^{\frac{\alpha}{2(\alpha-1)}}{\rm exp}\left[({\alpha-1})\left(\frac{i\beta y_j}{\alpha}\right)^{\frac{\alpha}{\alpha-1}}\right]\Bigg\}\,.
\end{align}
Thus, from the recurrence relations, we conclude that
\begin{align}\label{eq:calEnsmallb}
{\cal E}_{n}[\beta y_j\ll1;\alpha]=  \frac{1}{\alpha}\sum_{m=0}^\infty \frac{\Gamma \left[\frac{1+m+n}{\alpha }\right]}{m!}(i\beta y_j)^m\,,
\end{align}
and, at leading order, that
\begin{align}\label{eq:calEnlargeb}
{\cal E}_{n}[\beta y_j\gg1;\alpha]=&\frac{1}{(iy_j)^{n}}\frac{\partial^{n}}{\partial\beta^{n}}{\cal E}_{0}[\beta y_j;\alpha]\approx\frac{i\, n!}{(-i\beta y_j)^{1+n}}\left(1-\frac{\Gamma[n+\alpha+1]}{(-i\beta y_j)^{\alpha}}\right)\nonumber\\&+\frac{1}{i\beta y_j}
\sqrt{\frac{2\pi\alpha}{(\alpha-1)}}\left(\frac{i\beta y_j}{\alpha}\right)^{\frac{2n+\alpha}{2(\alpha-1)}}{\rm exp}\left[({\alpha-1})\left(\frac{i\beta y_j}{\alpha}\right)^{\frac{\alpha}{\alpha-1}}\right]+...\,.
\end{align}
Note that we can find more precise expressions for ${\cal E}_{n}[\beta y_j\gg1;\alpha]$ by considering all terms that follow from inserting Eq.~\eqref{eq:calE0largeb} into \eqref{eq:calEnrecurrence}.

We can also improve the expansions of ${\rm Sei}$ and ${\rm cei}$ using Eq.~\eqref{eq:calEntoEm1}. First for $\beta y_j\ll1$ we find that
\begin{align}\label{eq:ceiseiapplowb}
i\int_0^{\beta y_j\ll1} dt\,{\cal E}_{0}[t;\alpha]=-{\rm cei}[\beta y_j]+i\,{\rm Sei}[\beta y_j]= \frac{1}{\alpha}\sum_{m=0}^\infty \frac{\Gamma \left[\frac{1+m}{\alpha }\right]}{(m+1)!}(i\beta y_j)^{m+1}\,.
\end{align}
For large $\beta y_j$ Eq.~\eqref{eq:calEntoEm1} does not apply directly. Instead, we take the indefinite integral and fix the constant by matching the known asymptotic behaviours. Namely, we consider that
\begin{align}
i\int d(\beta y_j)\,{\cal E}_{0}[\beta y_j;\alpha]=-{\rm cei}[\beta y_j]+i\,{\rm Sei}[\beta y_j]+C\,,
\end{align} 
and fix $C=\gamma_E  \left(1-{\alpha }^{-1}\right)+i\pi{\rm sign}[y_j](\sqrt{2/\alpha}-1/{2})$ by using Eqs.~\eqref{eq:Seiasympt} and \eqref{eq:ceiasympt}. Then, we have that for $\beta y_j\gg1$
\begin{align}\label{eq:ceiseilargeb}
-{\rm cei}[\beta y_j]+i{\rm Sei}[\beta y_j]\approx-\ln|\beta y_j|+&{i \pi }\,{\rm sign}[y_j]\left(\frac{1}{2}-\sqrt{\frac{2}{\alpha}}\right)-\gamma_E  \left(1-{\alpha }^{-1}\right)\nonumber\\&-\frac{\Gamma[\alpha]}{(-i \beta y_j)^{\alpha }}-i\pi\sqrt{\frac{2 
   }{\alpha}}\text{Erf}\left[i\sqrt{\alpha-1 } \left(\frac{i \beta y_j}{\alpha
   }\right)^{\frac{\alpha }{2 (\alpha -1)}}\right]\,.
\end{align}

\subsubsection{Accurate approximations for the integrals}
It is interesting to note that the expansion for $\beta y_j\ll1$ \eqref{eq:calEnsmallb} can be pushed to the regime $\beta y_j\gg1$ for large enough $m$. This is because for $\alpha>1$ the factorial grows faster than the Gamma function for large enough $m$. Since we know that the expansion for $\beta y_j\gg1$ is quite good already for $\beta y_j>\alpha$, we can match both at $\beta y_j=c\alpha$ if we cut the expansion \eqref{eq:calEnsmallb} at $m>m_{\rm cut}$ where $m_{\rm cut}$ is a solution to
\begin{align}
e^{\frac{(\alpha -1) m_{\rm cut}}{\alpha }} m_{\rm cut}^{\frac{m_{\rm cut}+2 n+2}{\alpha }-m_{\rm cut}-2} \alpha ^{-\frac{-\alpha
   +m_{\rm cut}+2 n+2}{\alpha }}(c \,\alpha)^{m_{\rm cut}}=1\,.
\end{align}
For $c=2$ and $\alpha\leq 7$ and $n\leq 5$, again the relevant choice of parameters for our study, we find that with $m_{\rm cut}\sim 40-50$ we can use the low $\beta y_j$ expansion \eqref{eq:calEnsmallb} until $\beta y_j=2\alpha$ where we match with the large $\beta y_j$ expansion \eqref{eq:calEnlargeb}. This approach is faster than doing each integral numerically. An even more efficient approach would be to use numerical interpolation to bridge the low $\beta y_j$ with the high $\beta y_j$ regime to avoid using large Taylor expansions. For our purposes, the large Taylor expansions are enough though.

\subsection{Properties of the integral \texorpdfstring{${E}_n$}{}}

We now turn our attention to the integral $E_n$ \eqref{eq:baseE}. For $E_n$, we find the following recurrence relations. First, by taking one derivative with respect to $\lambda_j$ we find
\begin{align}\label{eq:partiallambdaE}
\frac{\partial}{\partial\lambda}E_n[\lambda_j;\alpha]=iE_{n+1}[\lambda_j;\alpha]-E_{n+\alpha}[\lambda_j;\alpha]\,.
\end{align}
Then, by integration by parts one has that, for $n>0$, 
\begin{align}
E_n[\lambda_j;\alpha]=\frac{\lambda_j}{n+1}\left(\alpha \,E_{n+\alpha}[\lambda_j;\alpha]-iE_{n+1}[\lambda_j;\alpha]\right)\,.
\end{align}
Combining the two to eliminate $E_{n+\alpha}[\lambda_j;\alpha]$ yields, for $n>-1$, 
\begin{align}
E_{n+1}[\lambda_j;\alpha]=-i\frac{\alpha}{\alpha-1}\left(\frac{\partial}{\partial\lambda_j}E_n[\lambda_j;\alpha]+\frac{n+1}{\lambda_j\alpha}E_{n}[\lambda_j;\alpha]\right)\,.
\end{align}
Thus, if we know $E_0[\lambda_j;\alpha]$ analytically, then we know all $E_n[\lambda_j;\alpha]$ with $n>0$. We also note that
\begin{align}
\alpha E_{\alpha-1}[\lambda_j;\alpha]=iE_0[\lambda_j;\alpha]+\frac{1}{\lambda_j}\,.
\end{align}
We can use this expression and Eq.~\eqref{eq:partiallambdaE} to write
\begin{align}
\frac{\partial}{\partial\lambda_j}E_{-1}[\lambda_j;\alpha]=-\frac{1}{\lambda_j\alpha}+i\frac{\alpha-1}{\alpha}E_{0}[\lambda_j;\alpha]\,.
\end{align}

\subsection{Asymptotic behaviors of the integral \texorpdfstring{${E}_n$}{} \label{app:asymptoticEn}}
We now turn to the asymptotic behaviours of the integral $E_n$ \eqref{eq:baseE}. As explained above, we shall focus only on $E_0[\lambda_j;\alpha]$. The rest follow from the recurrence relations.

First, we focus on the limit $\lambda_j\ll1$. In that case, most of the contribution from the integral comes from the regime where $z^\alpha\gg z$ for $\alpha>1$. Taylor expanding the imaginary part and integrating gives
\begin{align}
E_{0}[\lambda_j\ll1;\alpha]\approx \frac{1}{\alpha}\sum_{m=0}^\infty  \frac{i^m\lambda_j^{m-\frac{1+m}{\alpha}}}{m!}\Gamma\left[\frac{1+m+n}{\alpha}\right]\,,
\end{align}
where $\Gamma[x]$ is the Gamma function.

Second, let us focus on the regime where $\lambda_j\gg 1$ and the integrand is highly oscillating. In that case, we find asymptotic approximations by extending the integral to the complex plane and using the steepest descent approximation. The number of saddle points given by a stationary phase depends on the value of $\alpha$. For $\alpha>1$, the saddle points are located at $\alpha z_s^{\alpha-1}-i=0$. Writing $z_s=R\,e^{i\pi\theta}$, we find that $R=\alpha^{\frac{-1}{\alpha-1}}$ and $\theta=\frac{1+4m}{2(\alpha-1)}$ where $m\in\mathbb{Z}$ and $m\in[{\rm Ceiling}[\frac{1-2\alpha}{4}],{\rm Floor}[\frac{2\alpha-3}{4}]]$. For example, for $\alpha=2$, $m=0$ and $\theta=1$. For $\alpha=3$, $m=\left\{-1,0\right\}$ and $\theta=\left\{1/4,-3/4\right\}$. For $\alpha=4$, $m=\left\{-1,0,1\right\}$ and $\theta=\left\{-1/2,1/6,5/6\right\}$, etcetera. However, there is always a saddle point closer to the positive real line with $m=0$ and $\theta_0=\frac{1}{2(\alpha-1)}$.\footnote{For $\alpha=2$ the only saddle point exactly falls at $z_s=i$ but this case can be carried out analytically.} 

With the above information, we deform the contour as follows. We first integrate over the imaginary axis from $0$ to $z_c<i R$. We then close the contour by passing by the saddle close to the real axis, let us call it $z_{s,0}=\alpha^{\frac{-1}{\alpha-1}}\,e^{\frac{i\pi}{2(\alpha-1)}}$. So for $\lambda_j\gg1$ we have that
\begin{align}
 \int_{0}^\infty + \int_{0}^{z_c<iR}+\int_{\cal C_{\rm saddle}} dz\,e^{-\lambda_j\left({z^\alpha}-iz\right)}\approx 0\,.
 \end{align} 
 The integration over the imaginary axis is approximately given by
\begin{align}
i\int_{0}^{iR} {dz} \,e^{-\lambda_j\left(z+i^\alpha{z^\alpha}\right)}\approx \frac{i}{\lambda_j}+\frac{\alpha\Gamma[\alpha]}{\lambda_j^\alpha} {\rm e}^{i(\alpha-1)\frac{\pi}{2}}+{\cal O}(e^{-\lambda_jR})\,,
\end{align}
where we neglect exponentially suppressed terms near $z\sim i\,R$.
The contour through the saddle point yields
\begin{align}
\int_{\cal C_{\rm saddle}} dz\,e^{-\lambda_j\left({z^\alpha}-iz\right)}\approx i\sqrt{\frac{2\pi|z_s|}{\lambda(\alpha-1)}}e^{i\lambda z_s\frac{\alpha-1}{\alpha}}
\end{align}
Therefore, we find for $\lambda_j\gg1$ that
\begin{align}
E_{0}[\lambda_j\gg1;\alpha]\approx \frac{i}{\lambda_j}+\frac{\alpha}{\lambda_j^\alpha} {\rm e}^{i(\alpha-1)\frac{\pi}{2}}+\sqrt{\frac{2\pi|z_s|}{\lambda_j(\alpha-1)}}{\rm exp}\left[i\lambda_j z_s\frac{\alpha-1}{\alpha}-i\frac{\pi}{4}\frac{\alpha-2}{\alpha-1}\right]\,.
\end{align}

\subsection{Relations of the generalized cosine integral}

For the ${\rm Cei}^0_1[\beta_i]$ function it always appears as a difference, so it is sometimes more convenient to work with two related functions, namely
\begin{align}
{\rm cei}[\beta y_j]=\int_{0}^\infty dz\,\frac{e^{-z^\alpha}}{z} [1-\cos (\beta y_j z)]\,,
\end{align}
and
\begin{align}
{\rm ceic}[\beta y_j]=\int_{0}^\infty dz\,\frac{\cos (\beta y_j z)}{z} (1-e^{-z^\alpha})\,,
\end{align}
which are regular and can be treated separately. These functions are related by
\begin{align}
  {\rm cei}[\beta y_j]= {\rm ceic}[\beta y_j]+\ln|\beta y_j|+\left(1-\alpha^{-1}\right)\gamma_E\,,
\end{align}
where $\gamma_E\approx 0.577$. In the regime $\beta y_j\gg1$, we have that ${\rm ceic}[\beta y_j]\sim 0$ because the high-frequency oscillations average out and so we conclude that ${\rm cei}[\beta y_j]\sim \ln|\beta y_j|+\left(1-\alpha^{-1}\right)\gamma_E$.

\section{Details of the calculations of the Kernel\label{app:detailskernel}}
In this appendix, we provide details on several steps in the calculation of the induced GW kernel. In particular, we give the exact expressions for the terms in Eq.~\eqref{eq:IjIygamma}. To do so, we first define for compactness the generalized cosine and sine exponential integrals as
\begin{align}\label{eq:Ceigeneral}
{\rm Cei}^m_n[y_i]=\int_{\epsilon}^\infty \frac{dx}{x^n}\left(F_{\cal X}[{\cal X}]\right)^m e^{-{(v^2+u^2)\kappa_{D\star}^2}F[{\cal X}]} \cos (y_ix)\,,
\end{align}
and
\begin{align}\label{eq:Seigeneral}
{\rm Sei}^m_n[y_i]=\int_{\epsilon}^\infty \frac{dx}{x^n}\left(F_{\cal X}[{\cal X}]\right)^m e^{-{(v^2+u^2)\kappa_{D\star}^2}F[{\cal X}]} \sin (y_ix)\,,
\end{align}
where ${\cal X}\equiv x/x_*=\kappa x$. Note that we place a cut-off at the lower integration limit because the integrals, as defined, do not converge when $\epsilon\to 0$. However, the whole expression of the Kernel converges. Thus, one should take the limit $\epsilon\to 0$ at the end, namely
\begin{align}\label{eq:fullIyIj}
I_{y}(k,\tau,u,v)=\lim_{\epsilon\to 0}\left\{I^{(0)}_y+\gamma_*^2 I^{(1)}_y+\gamma_*^4 I^{(2)}_y\right\}\quad,\quad I_{j}(x,u,v)=\lim_{\epsilon\to 0}\left\{I^{(0)}_j+\gamma_*^2 I^{(1)}_j+\gamma_*^4 I^{(2)}_j\right\}\,.
\end{align}
Also, note that in the main text, we defined
\begin{align}
{\rm Cei}[y_i]\equiv{\rm Cei}^0_1[y_i]\quad{\rm and}\quad {\rm Sei}[y_i]\equiv{\rm Sei}^0_1[y_i]\,,
\end{align}
for simplicity. Furthermore, these functions reduce to the standard cosine and sine integrals, ${\rm Ci}[x]$ and ${\rm Si}[x]$, when $F=0$.

The explicit expression for the various terms in Eq.~\eqref{eq:IjIygamma} are given by
\begin{align}
I^{(0)}_y=&\frac{1}{c_s^3 u^2 v^2}\left[(u-v)\left({\rm Cei}^0_2[y_1]-{\rm Cei}^0_2[y_2]\right)+(u+v) \left({\rm Cei}^0_2[y_3]-{\rm Cei}^0_2[y_4]\right)\right]\nonumber\\&
   +\frac{3}{c_s^5 u^3 v^3} \left[(u-v)\left( {\rm Cei}^0_4[y_1]- {\rm Cei}^0_4[y_2]\right)-(u+v)\left({\rm Cei}^0_4[y_3]-{\rm Cei}^0_4[y_4]\right)\right]\nonumber\\&
   +\frac{1}{2c_s^2 u v}\left[{\rm Sei}^0_1[y_1]+{\rm Sei}^0_1[y_2]-{\rm Sei}^0_1[y_3]-{\rm Sei}^0_1[y_4]\right]\nonumber\\&
   -\frac{1}{c_s^4 u^3 v^3}\left[\left(u^2-3 u v+v^2\right) \left({\rm Sei}^0_3[y_1]+
   {\rm Sei}^0_3[y_2]\right)-\left(u^2+3 u v+v^2\right) \left({\rm Sei}^0_3[y_3]+{\rm Sei}^0_3[y_4]\right)\right]\nonumber\\&
   +\frac{3}{c_s^6 u^3 v^3} \left[{\rm Sei}^0_5[y_1]+{\rm Sei}^0_5[y_2]-{\rm Sei}^0_5[y_3]-{\rm Sei}^0_5[y_4]\right]\,,
\end{align}
\begin{align}
I^{(1)}_y=&\frac{1}{2 c_s^3 u v}\left[-(u-v) \left({\rm Cei}^1_1[y_1]- {\rm Cei}^1_1[y_2]\right)+(u+v) \left({\rm Cei}^1_1[y_3]-{\rm Cei}^1_1[y_4]\right)\right]\nonumber\\&
   +\frac{u^2+v^2}{ c_s^5
   u^3 v^3}\left[ (u-v) \left({\rm Cei}^1_3[y_1]- {\rm Cei}^1_3[y_2]\right)-(u+v) \left({\rm Cei}^1_3[y_3]-{\rm Cei}^1_3[y_4]\right)\right]\nonumber\\&
   +\frac{1}{ c_s^4 u^2 v^2}\left[\left(u^2-u v+v^2\right) \left({\rm Sei}^1_2[y_1]+
   {\rm Sei}^1_2[y_2]\right)+\left(u^2+u v+v^2\right) \left({\rm Sei}^1_2[y_3]+{\rm Sei}^1_2[y_4]\right)\right]\nonumber\\&
   +\frac{u^2+v^2}{ c_s^6 u^3 v^3}\left[ {\rm Sei}^1_4[y_1]+{\rm Sei}^1_4[y_2]-{\rm Sei}^1_4[y_3]-{\rm Sei}^1_4[y_4]\right]\,,
\end{align}
\begin{align}
I^{(2)}_y=&\frac{1}{2 c_s^5 u v}\left[(u-v) \left({\rm Cei}^2_2[y_1]- {\rm Cei}^2_2[y_2]\right)-(u+v) \left({\rm Cei}^2_2[y_3]-{\rm Cei}^2_2[y_4]\right)\right]+\nonumber\\&
   +\frac{1}{2 c_s^4}\left[{\rm Sei}^2_1[y_1]+{\rm Sei}^2_1[y_2]+{\rm Sei}^2_1[y_3]+{\rm Sei}^2_1[y_4]\right]\nonumber\\&
   +\frac{1}{2 c_s^6 u v}\left[{\rm Sei}^2_3[y_1]+{\rm Sei}^2_3[y_2]-{\rm Sei}^2_3[y_3]-{\rm Sei}^2_3[y_4]\right]\,,
\end{align}
\begin{align}
I^{(0)}_j=&\frac{1}{c_s^3 u^2 v^2}\left[(u-v)\left({\rm Sei}^0_2[y_1]-{\rm Sei}^0_2[y_2]\right)+(u+v) \left({\rm Sei}^0_2[y_3]-{\rm Sei}^0_2[y_4]\right)\right]\nonumber\\&
   +\frac{3}{c_s^5 u^3 v^3} \left[(u-v)\left( {\rm Sei}^0_4[y_1]- {\rm Sei}^0_4[y_2]\right)-(u+v)\left({\rm Sei}^0_4[y_3]-{\rm Sei}^0_4[y_4]\right)\right]\nonumber\\&
   -\frac{1}{2c_s^2 u v}\left[{\rm Cei}^0_1[y_1]+{\rm Cei}^0_1[y_2]-{\rm Cei}^0_1[y_3]-{\rm Cei}^0_1[y_4]\right]\nonumber\\&
   +\frac{1}{c_s^4 u^3 v^3}\left[\left(u^2-3 u v+v^2\right) \left({\rm Cei}^0_3[y_1]+
   {\rm Cei}^0_3[y_2]\right)-\left(u^2+3 u v+v^2\right) \left({\rm Cei}^0_3[y_3]+{\rm Cei}^0_3[y_4]\right)\right]\nonumber\\&
   -\frac{3}{c_s^6 u^3 v^3} \left[{\rm Cei}^0_5[y_1]+{\rm Cei}^0_5[y_2]-{\rm Cei}^0_5[y_3]-{\rm Cei}^0_5[y_4]\right]\,,
\end{align}
\begin{align}
I^{(1)}_j=&\frac{1}{2 c_s^3 u v}\left[-(u-v) \left({\rm Sei}^1_1[y_1]- {\rm Sei}^1_1[y_2]\right)+(u+v) \left({\rm Sei}^1_1[y_3]-{\rm Sei}^1_1[y_4]\right)\right]\nonumber\\&
   +\frac{u^2+v^2}{ c_s^5
   u^3 v^3}\left[ (u-v) \left({\rm Sei}^1_3[y_1]- {\rm Sei}^1_3[y_2]\right)-(u+v) \left({\rm Sei}^1_3[y_3]-{\rm Sei}^1_3[y_4]\right)\right]\nonumber\\&
   -\frac{1}{ c_s^4 u^2 v^2}\left[\left(u^2-u v+v^2\right) \left({\rm Cei}^1_2[y_1]+
   {\rm Cei}^1_2[y_2]\right)+\left(u^2+u v+v^2\right) \left({\rm Cei}^1_2[y_3]+{\rm Cei}^1_2[y_4]\right)\right]\nonumber\\&
   -\frac{u^2+v^2}{ c_s^6 u^3 v^3}\left[ {\rm Cei}^1_4[y_1]+{\rm Cei}^1_4[y_2]-{\rm Cei}^1_4[y_3]-{\rm Cei}^1_4[y_4]\right]\,,
\end{align}
\begin{align}
I^{(2)}_j=&\frac{1}{2 c_s^5 u v}\left[(u-v) \left({\rm Sei}^2_2[y_1]- {\rm Sei}^2_2[y_2]\right)-(u+v) \left({\rm Sei}^2_2[y_3]-{\rm Sei}^2_2[y_4]\right)\right]+\nonumber\\&
   -\frac{1}{2c_s^4}\left[{\rm Cei}^2_1[y_1]+{\rm Cei}^2_1[y_2]+{\rm Cei}^2_1[y_3]+{\rm Cei}^2_1[y_4]\right]\nonumber\\&
   -\frac{1}{2 c_s^6 u v}\left[{\rm Cei}^2_3[y_1]+{\rm Cei}^2_3[y_2]-{\rm Cei}^2_3[y_3]-{\rm Cei}^2_3[y_4]\right]\,.
\end{align}

Now, we can simplify the terms $I^{(0)}_j$ and $I^{(0)}_y$ in Eq.~\eqref{eq:fullIyIj} by integrating by parts using the following relations,
\begin{align}
   (n-1){\rm Sei}^0_n[y_i x]=&-\frac{
   \sin[y_i x] e^{-\kappa_{D\star}^2 \left(u^2+v^2\right)
   F[X]}}{\epsilon^{n-1}}
   +y_i {\rm Cei}_{n-1}^0[y_i x]-\gamma_* ^2
   \left(u^2+v^2\right){\rm Sei}^1_{n-1}[y_i x]\,,
\end{align}
and
\begin{align}
   (n-1){\rm Cei}^0_n[y_i x]=&-\frac{
   \cos[y_i x] e^{-\kappa_{D\star}^2 \left(u^2+v^2\right)
   F[X]}}{\epsilon^{n-1}}
   -y_i {\rm Sei}_{n-1}^0[y_i x]-\gamma_* ^2
   \left(u^2+v^2\right){\rm Cei}^1_{n-1}[y_i x]\,.
\end{align}
In doing so, we obtain a new expression for the kernel given by
\begin{align}
I_{y}(x,u,v)&=\lim_{\epsilon\to 0}\left\{\tilde I^0_y+\gamma_*^2\left(\tilde I^{(1)}_y+ I^{(1)}_y\right)+\gamma_*^4 I^{(2)}_y\right\}\,,\nonumber\\
I_{j}(x,u,v)&=\lim_{\epsilon\to 0}\left\{\tilde I^0_j+\gamma_*^2\left(\tilde I^{(1)}_j+I^{(1)}_j\right)+\gamma_*^4 I^{(2)}_j\right\}\,,
\end{align}
where we denote the new terms which result after the integration by parts with a tilde, namely $\tilde I^{(0)}_y$, $\tilde I^{(0)}_j$ as well as $\tilde I^{(1)}_y$ and $\tilde I^{(1)}_j$. Their explicit expression reads
\begin{align}
\tilde I^{(0)}_y=&
   \frac{\left(1-c_s^2 \left(u^2+v^2\right)\right)^2}{8 c_s^6 u^3 v^3}\left[{\rm Sei}^0_1[y_1]+{\rm Sei}^0_1[y_2]-{\rm Sei}^0_1[y_3]-{\rm Sei}^0_1[y_4]\right]\,,
\end{align}

\begin{align}
\tilde I^{(0)}_j=&-\frac{1-c_s^2 \left(u^2+v^2\right)}{2 c_s^4 u^2 v^2}
   -\frac{\left(1-c_s^2 \left(u^2+v^2\right)\right)^2}{8 c_s^6 u^3 v^3}\left[{\rm Cei}^0_1[y_1]+{\rm Cei}^0_1[y_2]-{\rm Cei}^0_1[y_3]-{\rm Cei}^0_1[y_4]\right]\,,
\end{align}
and
\begin{align}
\tilde I^{(1)}_y=&\frac{(u^2+v^2)}{8  c_s^6  u^3  v^3}\Bigg\{-2(1+3 c_s (u-v)) {\rm Cei}^1_3[y_1]-2(1-3 c_s (u-v)) {\rm Cei}^1_3[y_2]\nonumber\\&+2(1+3 c_s
   (u+v)) {\rm Cei}^1_3[y_3]+2(1-3 c_s (u+v)) {\rm Cei}^1_3[y_4]\nonumber\\&+y_1y_2 y_3 y_4\left[
   \frac{{\rm Cei}^1_1[y_1]}{y_1}+\frac{{\rm Cei}^1_1[y_2]}{y_2}-\frac{
   {\rm Cei}^1_1[y_3]}{y_3}-\frac{{\rm Cei}^1_1[y_4]}{y_4}\right]\nonumber\\&
   +\left(1-2 c_s (u-v)+c_s^2 \left(u^2-6
   u v+v^2\right)\right) {\rm Sei}^1_2[y_2]\nonumber\\&-\left(1+2 c_s (u+v)+c_s^2 \left(u^2+6 u
   v+v^2\right)\right) {\rm Sei}^1_2[y_3]\nonumber\\&-\left(1-2 c_s (u+v)+c_s^2 \left(u^2+6 u
   v+v^2\right)\right) {\rm Sei}^1_2[y_4]\nonumber\\&+\left(1+2 c_s (u-v)+c_s^2 \left(u^2-6 u v+v^2\right)\right) {\rm Sei}^1_2[y_1]\nonumber\\&
   -6\left[ {\rm Sei}^1_4[y_1]+{\rm Sei}^1_4[y_2]- {\rm Sei}^1_4[y_3]-{\rm Sei}^1_4[y_4]\right]\Bigg\}\,,
\end{align}

\begin{align}
\tilde I^{(1)}_j=&\frac{(u^2+v^2)}{8  c_s^6  u^3  v^3}\Bigg\{-\left(1+2 c_s (u-v)+c_s^2 \left(u^2-6 u v+v^2\right)\right){\rm Cei}^1_2[y_1]\nonumber\\&-\left(1-2 c_s (u-v)+c_s^2 \left(u^2-6 u v+v^2\right)\right) {\rm Cei}^1_2[y_2]\nonumber\\&
+\left(1+2 c_s (u+v)+c_s^2 \left(u^2+6 u v+v^2\right)\right) {\rm Cei}^1_2[y_3]\nonumber\\&+\left(1-2 c_s (u+v)+c_s^2 \left(u^2+6 u v+v^2\right)\right) {\rm Cei}^1_2[y_4]\nonumber\\&
+6\left[
   {\rm Cei}^1_4[y_1]+ {\rm Cei}^1_4[y_2]- {\rm Cei}^1_4[y_3]- {\rm Cei}^1_4[y_4]\right]\nonumber\\&
   -2(1+3c_s (u-v)) {\rm Sei}^1_3[y_1]-2(1-3 c_s (u-v)) {\rm Sei}^1_3[y_2]\nonumber\\&
   +2(1+3 c_s(u+v)) {\rm Sei}^1_3[y_3]+2(1-3 c_s (u+v)) {\rm Sei}^1_3[y_4]\nonumber\\&
   +y_1y_2 y_3 y_4\left[
   \frac{{\rm Sei}^1_1[y_1]}{y_1}+ \frac{{\rm Sei}^1_1[y_2]}{y_2}-\frac{
   {\rm Sei}^1_1[y_3]}{y_3}-\frac{{\rm Sei}^1_1[y_4]}{y_4}\right]\Bigg\}\,.
\end{align}

\section{Exact expressions for the Induced GW Kernel\label{app:otherpowerlaw}}

In the cases where the damping scale is a power-law given by Eq.~\eqref{eq:F} with $\alpha=1$ and $\alpha=2$ we find exact analytical expressions for the total kernel \eqref{eq:calIjIy}. We provide them explicitly in this appendix.

\subsection{The case of \texorpdfstring{$\alpha=1$}{}}

For $\alpha=1$ we have that $\beta$ \eqref{eq:betaj} reads
\begin{align}\label{eq:betaja1}
\beta_1 =\frac{1}{\gamma_*^2\kappa(u^2+v^2)}\,,
\end{align}
where in the last step we used that $\gamma_*=\kappa_{D}/\kappa$ from Eq.~\eqref{eq:gamma}. The sum \eqref{eq:calIjIy} is now given by
\begin{align}\label{eq:calIjIya1}
I_j={\cal I}^{(0)}_j+\sum_{m=1}^{4} \gamma_*^{2m}{\cal I}^{(m)}_j\quad,\quad I_y={\cal I}^{(0)}_y+\sum_{m=1}^{4} \gamma_*^{2m}{\cal I}^{(m)}_y\,,
\end{align}
We provide the explicit expression of each individual term below. Note that $\gamma_*$ will always appear in the combination $\gamma_*^2\kappa$, which could be traded for $\beta$. We will not do so here. First, we have that
\begin{align}\label{eq:calIj0a1}
{\cal I}^{(0)}_j=&-\frac{1-c_s^2 \left(u^2+v^2\right)}{2 c_s^4 u^2 v^2}\left(1
   -\frac{1-c_s^2 \left(u^2+v^2\right)}{8 c_s^2 u v}L[\beta_1,u,v]\right)\,,
\end{align}
and
\begin{align}\label{eq:calIy0a1}
{\cal I}^{(0)}_y= &
   \frac{\left(1-c_s^2 \left(u^2+v^2\right)\right)^2}{8 c_s^6 u^3 v^3}A[\beta_1,u,v]\,,
\end{align}
where for compactness and later use we have defined
\begin{align}
L[\beta_1,u,v]\equiv \ln\left[\frac{(1+\beta_1 ^2y_1^2)(1+\beta_1 ^2y_2^2)}{(1+\beta_1 ^2y_3^2)(1+\beta_1 ^2y_4^2)}\right]\,,
\end{align}
and
\begin{align}
A[\beta_1,u,v]\equiv \tan^{-1}[\beta_1 y_1]+\tan^{-1}[\beta_1 y_2]-\tan^{-1}[\beta_1 y_3]-\tan^{-1}[\beta_1 y_4]\,.
\end{align}
The other terms read
\begin{align}
\kappa^{-1}{\cal I}^{(1)}_j&=\frac{u^2+v^2-3 c_s^2 u^2 v^2}{3 c_s^6 u^3 v^3}A[\beta_1,u,v]\nonumber\\&+\frac{1}{6 c_s^3 u^3 v^3}\Big(-\left(2 u^5-u^3 v^2+u^2 v^3-2 v^5\right)
   (\tan^{-1}[\beta_1 y_1]-\tan^{-1}[\beta_1 y_2])\nonumber\\&\qquad\qquad\qquad+\left(2 u^5-u^3 v^2-u^2 v^3+2 v^5\right)
   (\tan^{-1}[\beta_1 y_3]-\tan^{-1}[\beta_1 y_4]))\Big)\,,\\
\kappa^{-2} {\cal I}^{(2)}_j&=-\frac{\left(u^4-4 u^2 v^2+v^4\right)}{6 c_s^4 u^2
   v^2}-\frac{u^4+3 u^2 v^2+v^4+c_s^2 \left(u^6-2 u^4 v^2-2 u^2 v^4+v^6\right)}{8c_s^6 u^3 v^3}L[\beta_1,u,v]\,,\\
\kappa^{-3}{\cal I}^{(3)}_j&=-\frac{\left(u^2+v^2\right)}{2 c_s^6 u v}A[\beta_1,u,v]\,,\\
\kappa^{-4}{\cal I}^{(4)}_j&=-\frac{\left(u^2+v^2\right)^2 \left(u^4-4 u^2
   v^2+v^4\right)}{48 c_s^6 u^3 v^3}L[\beta_1,u,v]\,,
\end{align}
and
\begin{align}
\kappa^{-1}{\cal I}^{(1)}_y&=\frac{u^2+v^2}{3c_s^4 u^2 v^2}+\frac{u^2+v^2-3 c_s^2 u^2 v^2}{6 c_s^6 u^3 v^3}L[\beta_1,u,v]\nonumber\\&+\frac{1}{12 c_s^3 u^3 v^3}\Big(\left(2 u^5-u^3 v^2+u^2 v^3-2 v^5\right)
   (\ln\left(1+\beta_1^2y_1^2\right)-\ln\left(1+\beta_1^2y_2^2\right))\nonumber\\&-\left(2 u^5-u^3 v^2-u^2 v^3+2 v^5\right)
   (\ln\left(1+\beta_1^2y_3^2\right)-\ln\left(1+\beta_1^2y_4^2\right))\Big)\,,\\
\kappa^{-2}{\cal I}^{(2)}_y&=-\frac{u^4+3 u^2 v^2+v^4+c_s^2 \left(u^6-2 u^4 v^2-2 u^2 v^4+v^6\right)}{4c_s^6 u^3 v^3}A[\beta_1,u,v]\,,\\
\kappa^{-3}{\cal I}^{(3)}_y&=\frac{\left(u^2+v^2\right)}{4 c_s^6 u v}L[\beta_1,u,v]\,,\\
\kappa^{-4}{\cal I}^{(4)}_y&=-\frac{\left(u^2+v^2\right)^2 \left(u^4-4 u^2
   v^2+v^4\right)}{24 c_s^6 u^3 v^3}A[\beta_1,u,v]\,.
\end{align}

\subsection{The case of \texorpdfstring{$\alpha=2$}{}}

When $\alpha=2$ we have that $\beta$ \eqref{eq:betaj} becomes $\kappa$ independent, namely
\begin{align}\label{eq:betaja2}
\beta_2 =\frac{1}{\gamma_*\sqrt{(u^2+v^2)}}\,.
\end{align}
The sum \eqref{eq:calIjIy} is now given by
\begin{align}\label{eq:calIjIya2}
I_j={\cal I}^{(0)}_j+\sum_{m=1}^{2} \gamma_*^{2m}{\cal I}^{(m)}_j\quad,\quad I_y={\cal I}^{(0)}_y+\sum_{m=1}^{2} \gamma_*^{2m}{\cal I}^{(m)}_y\,.
\end{align}
Let us spell each term separately. First, we find that
\begin{align}\label{eq:calIj0a2}
{\cal I}^{(0)}_j=&-\frac{1-c_s^2 \left(u^2+v^2\right)}{2 c_s^4 u^2 v^2}\left(1
   -\frac{1-c_s^2 \left(u^2+v^2\right)}{4 c_s^2 u v}D[\beta_2,u,v]\right)\,,
\end{align}
and
\begin{align}\label{eq:calIy0a2}
{\cal I}^{(0)}_y= &\pi
   \frac{\left(1-c_s^2 \left(u^2+v^2\right)\right)^2}{16 c_s^6 u^3 v^3}E[\beta_2,u,v]\,.
\end{align}
where we defined
\begin{align}
D[\beta_2,u,v]\equiv {\rm IDi}[\beta_2 y_1]+{\rm IDi}[\beta_2 y_2]-{\rm IDi}[\beta_2 y_3]-{\rm IDi}[\beta_2 y_4]\,,
\end{align}
with ${\rm IDi}[\beta y_1]$ being the integral of the Dawson function, given by Eq.~\eqref{eq:IDidef}, and
\begin{align}
E[\beta_2,u,v]={\rm Erf}\left[\frac{\beta_2 y_1}{2}\right]+{\rm Erf}\left[\frac{\beta_2 y_2}{2}\right]-{\rm Erf}\left[\frac{\beta_2 y_3}{2}\right]-{\rm Erf}\left[\frac{\beta_2 y_4}{2}\right]\,.
\end{align}

The other terms read
\begin{align}
{\cal I}^{(1)}_j=&\frac{u^2+v^2}{c_s^4 u^2 v^2}+\frac{u^2+v^2+c_s^2 \left(u^2-v^2\right)^2}{2 c_s^6 u^3 v^3}D[\beta_2,u,v]\nonumber\\&
-\frac{2\beta_2 }{c_s^3 u v}\left((u-v) \left({\rm Di}\left[\frac{\beta_2 y_1}{2}\right]-{\rm Di}\left[\frac{\beta_2 y_2}{2}\right]\right)-(u+v)
   \left({\rm Di}\left[\frac{\beta_2 y_3}{2}\right]-{\rm Di}\left[\frac{\beta_2 y_4}{2}\right]\right)\right)\nonumber\\&
   +\frac{\left(u^2+v^2\right)y_1y_2y_3y_4
   }{2 c_s^6 u^3 v^3}\beta_2 \left(\frac{1}{y_1}{\rm Di}\left[\frac{\beta_2 y_1}{2}\right]+\frac{1}{y_2}{\rm Di}\left[\frac{\beta_2 y_2}{2}\right]-\frac{1}{y_3}{\rm Di}\left[\frac{\beta_2 y_1}{2}\right]-\frac{1}{y_4}{\rm Di}\left[\frac{\beta_2 y_1}{2}\right]\right)\,,
\end{align}
\begin{align}
{\cal I}^{(2)}_j=&-\frac{16\beta_2^2}{c_s^4}-\frac{\left(u^2-v^2\right)^2}{2 c_s^6 u^3 v^3}D[\beta_2,u,v]-\frac{8}{c_s^4}\beta_2 ^2\left(2-\sum_{i=1}^4\beta_2 y_i{\rm Di}\left[\frac{\beta_2 y_i}{2}\right]\right)\nonumber\\&+\frac{\left(u^2+v^2\right)^2}{ c_s^6 u^3 v^3}\beta_2 \left({\rm Di}\left[\frac{\beta_2 y_1}{2}\right]+{\rm Di}\left[\frac{\beta_2 y_2}{2}\right]-{\rm Di}\left[\frac{\beta_2 y_3}{2}\right]-{\rm Di}\left[\frac{\beta_2 y_4}{2}\right]\right)\nonumber\\&
-\frac{\left(u^2-v^2\right)^2}{c_s^5 u^3 v^3}\beta_2 \left((u-v)
   \left({\rm Di}\left[\frac{\beta_2 y_1}{2}\right]-{\rm Di}\left[\frac{\beta_2 y_2}{2}\right]\right)-(u+v)
  \left({\rm Di}\left[\frac{\beta_2 y_3}{2}\right]-{\rm Di}\left[\frac{\beta_2 y_4}{2}\right]\right)\right)\,,
\end{align}
and
\begin{align}
{\cal I}^{(1)}_y&=\pi\frac{u^2+v^2+c_s^2 \left(u^2-v^2\right)^2}{4 c_s^6 u^3 v^3}E[\beta_2,u,v]\nonumber\\&
-\frac{\sqrt{\pi}\beta_2 }{4c_s^3 u v}\left((u-v) \left(e^{-(\beta_2 y_1)^2/4}-e^{-(\beta_2 y_2)^2/4}\right)-(u+v)
   \left(e^{-(\beta_2 y_3)^2/4}-e^{-(\beta_2 y_4)^2/4}\right)\right)\nonumber\\&
   +\frac{\left(u^2+v^2\right)y_1y_2y_3y_4
   }{16 c_s^6 u^3 v^3}\sqrt{\pi}\beta_2 \left(\frac{e^{-(\beta_2 y_1)^2/4}}{y_1}+\frac{e^{-(\beta_2 y_1)^2/4}}{y_2}-\frac{e^{-(\beta_2 y_1)^2/4}}{y_3}-\frac{e^{-(\beta_2 y_1)^2/4}}{y_4}\right)\,,
\end{align}
\begin{align}
{\cal I}^{(2)}_y&=-\frac{\left(u^2-v^2\right)^2}{4 c_s^6 u^3 v^3}E[\beta_2,u,v]+\frac{4\sqrt{\pi}}{c_s^4}\beta_2 ^3\sum_{i=1}^4 y_ie^{-(\beta_2 y_i)^2/4}\nonumber\\&
+\frac{\left(u^2+v^2\right)^2}{2 c_s^6 u^3 v^3}{\sqrt{\pi}\beta_2 }\left(e^{-(\beta_2 y_1)^2/4}+e^{-(\beta_2 y_2)^2/4}-e^{-(\beta_2 y_3)^2/4}-e^{-(\beta_2 y_4)^2/4}\right)\nonumber\\&
-\frac{\left(u^2-v^2\right)^2}{2 c_s^5 u^3 v^3}{\sqrt{\pi}\beta_2 }\left((u-v)
   \left(e^{-(\beta_2 y_1)^2/4}-e^{-(\beta_2 y_2)^2/4}\right)-(u+v)
  \left(e^{-(\beta_2 y_3)^2/4}-e^{-(\beta_2 y_4)^2/4}\right)\right)\,.
\end{align}

\bibliography{refgwscalar.bib}

\end{document}